\documentclass[aps,prb,showpacs,intlimits,amsmath,amssymb,twocolumn]{revtex4}

\usepackage{graphicx}
\usepackage{bm}

\DeclareMathOperator{\Img}{Im}
\DeclareMathOperator{\Real}{Re}
\DeclareMathOperator{\Tr}{Tr}

\begin{document}

\title{Optical and dc transport properties of a strongly correlated
charge density wave system:  exact solution in the ordered phase of the
spinless Falicov-Kimball model with dynamical mean-field theory}
\author{O.~P.~Matveev$^\dagger$, A.~M.~Shvaika$^\dagger$, and 
J.~K.~Freericks$^*$
}
\affiliation{$^\dagger$Institute for Condensed Matter Physics of the National Academy of Sciences of Ukraine,
Lviv, 79011 Ukraine}
\affiliation{$^*$Department of Physics, Georgetown University, Washington, DC
20057, U.S.A.}

\begin{abstract}
We derive the dynamical mean-field theory equations for transport in 
an ordered charge-density-wave phase on a bipartite lattice.  The formalism is
applied to the spinless Falicov-Kimball model on a hypercubic lattice
at half filling. We determine
the many-body density of states, the dc charge and heat conductivities,
and the optical conductivity.  Vertex corrections continue to vanish
within the ordered phase, but the density of states and the transport 
coefficients show anomalous behavior due to the rapid development of 
thermally activated subgap states.  We also examine the optical sum rule
and sum rules for the first three moments of the Green's functions 
within the ordered phase and see that the total optical spectral weight in the
ordered phase either decreases or increases depending on
the strength of the interactions.
\end{abstract}


\maketitle

\section{Introduction}
Dynamical mean-field theory was introduced almost two decades ago by
Brandt and Mielsch\cite{brandt_mielsch1}, who solved for the transition
temperature into a charge-density-wave (CDW) phase of the spinless
Falicov-Kimball model at half filling.  This work appeared shortly after
the idea of examining strongly correlated electrons in the limit of
infinite dimensions was introduced\cite{metzner_vollhardt}.  Since then,
the field of DMFT has emerged as one of the most powerful nonperturbative
techniques for solving the many-body problem.  While results for many
properties exist in the homogeneous (unordered) phase\cite{kotliar_review},
there has been little work in examining the properties of the ordered phase.
Brandt and Mielsch worked out the formalism for calculating ordered-phase
Green's functions\cite{brandt_mielsch2}, the order parameter was shown to
display anomalous behavior at weak coupling\cite{vandongen,chen_freericks},
and higher-period ordered phases have been examined on the Bethe 
lattice\cite{freericks_swiss}.  But, surprisingly, there has been no
work on the transport properties in the ordered phase. Indeed, it is
interesting to compare how transport varies in the homogeneous phase
versus the ordered phase.  At weak coupling, we anticipate the gap formation of
the CDW to greatly suppress the dc transport, while at strong coupling it may
be a much milder correction to the Mott-insulating behavior.  What is more
interesting is to examine the temperature dependence. For example, in systems
that are metallic at high temperature, the many-body DOS in the CDW phase
develops strong temperature dependence (with increasing $T$)
as the CDW gap region fills in due
to thermal excitations, until gap closure is complete at the transition
temperature.  But unlike the well-known superconducting case, where subgap
states tend not to form and the gap is simply reduced in size as $T$
increases, here we have a rapid development of subgap states, even though 
the CDW order parameter remains nonzero.  These subgap states should produce
anomalous behavior in the low-$T$ transport, and indeed we find this is
so but the quantitative behavior is not that different from exponential
activation of the transport. We anticipate our results should be relevant
to different experimental systems that display charge-density-wave order,
especially in compounds which are three-dimensional like\cite{cdw_exp}
BaBiO$_3$ and Ba$_{1-x}$K$_{x}$BiO$_3$.

This contribution is organized as follows:  In Section II, we present the
formalism for DMFT in the ordered phase including the techniques needed to
determine the optical conductivity and the dc transport.  We also determine
moment sum rules for the Green's functions in the ordered
phase. In Section III, we apply the formalism to numerical solutions of
the Falicov-Kimball model at half filling and show how the transport behaves
in the ordered phases.  Conclusions and a discussion follow in Section IV.

\section{Formalism for the ordered phase}

The Falicov-Kimball model\cite{falicov_kimball} was introduced in 1969 as
a model for metal-insulator transitions in rare-earth compounds and 
transition-metal oxides. The spinless version is arguably the simplest 
many-body problem that nevertheless possesses rich physics including
the Mott transition, order-disorder phase transitions, and phase
separation (for a review see Ref.~\onlinecite{freericks_review}). It involves
two kinds of electrons: mobile conduction electrons whose creation
and destruction operators are $\hat d_i^\dagger$ and $\hat d_i^{}$ 
at site $i$; and localized electrons whose creation and destruction 
operators are $\hat f_i^\dagger$ and $\hat f_i^{}$ at site $i$.  
The Falicov-Kimball Hamiltonian can be represented in terms of a local
operator and a hopping operator as follows
\begin{equation}
   \mathcal{\hat H}=
\sum_{i}\mathcal{\hat H}_{i}-\sum_{ij}t_{ij}\hat{d}_{i}^{\dag}\hat{d}_{j}^{},
\label{eq: ham_def}
\end{equation}
where $t_{ij}$ is the hopping matrix and
\begin{equation}
   \mathcal{\hat H}_{i}=U\hat{n}_{id}\hat{n}_{if}-\mu_{d}\hat{n}_{id}
-\mu_{f}\hat{n}_{if},
\label{eq: ham_def_local}
\end{equation}
is the local Hamiltonian with the number operators given by 
$\hat n_{id}=\hat d_i^\dagger\hat d_i^{}$ and
$\hat n_{if}=\hat f_i^\dagger\hat f_i^{}$.

If the lattice can be divided into two sublattices, and the hopping is nonzero
only between the two sublattices ({\it i.~e.}, there is no hopping within either
sublattice), then the lattice is called a bipartite lattice, and 
it has nesting at half filling in the noninteracting system, which implies
the fermi surface in the Brillouin zone has flat regions that are connected
by the zone-diagonal wavevector $Q=(\pi,\pi,\ldots )$. Nesting promotes
the formation of a CDW with the average filling of the electrons being uniform
on each sublattice, but changing from one sublattice to another.  This is
often called the checkerboard or chessboard CDW, and is the ordered phase
that we will examine in detail in this work.

\begin{figure}[htb]
   \raisebox{-.4\height}{\includegraphics[height=.15\textheight]{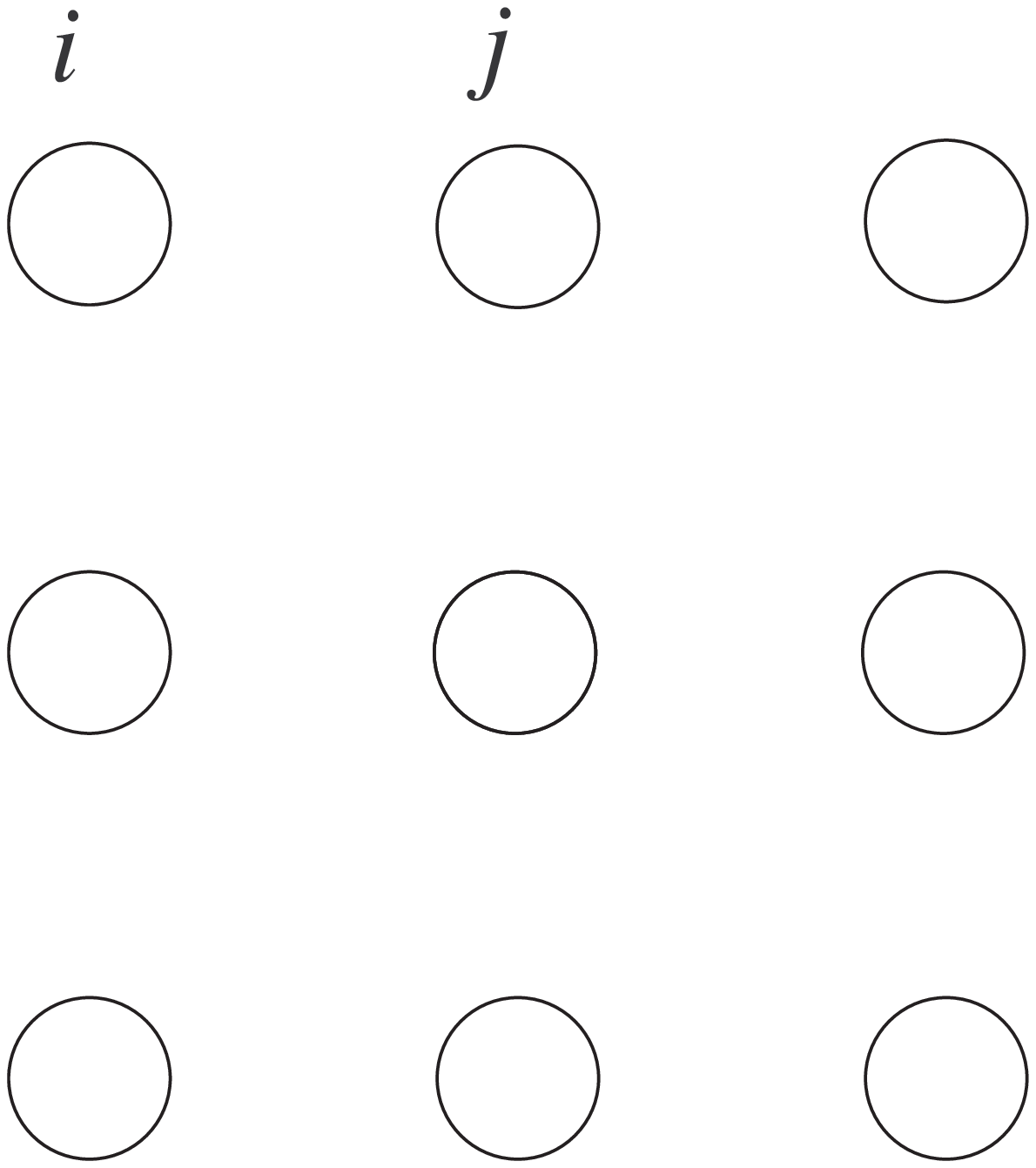}}
       \qquad$\to$\qquad
   \raisebox{-.4\height}{\includegraphics[height=.15\textheight]{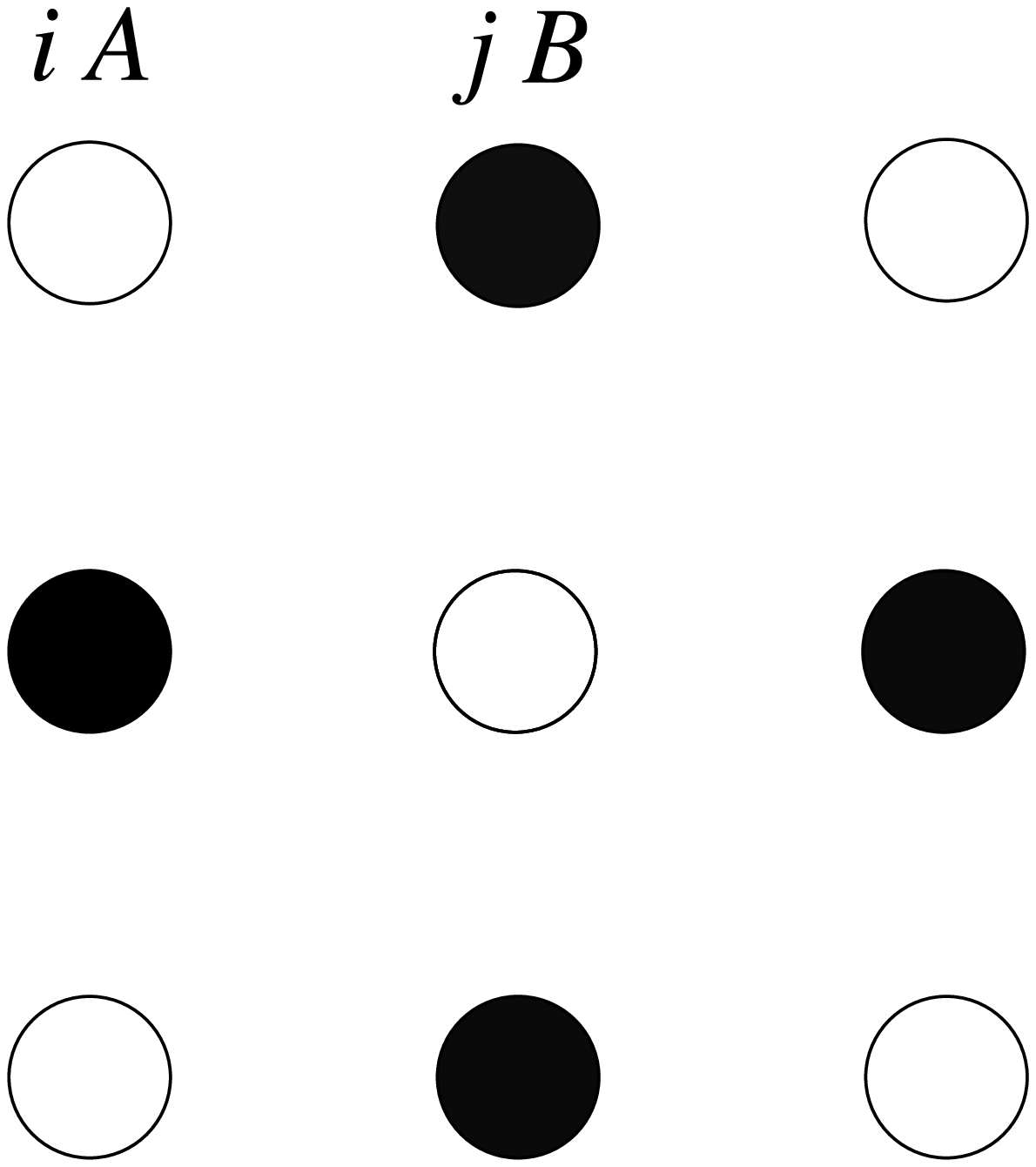}}
  \caption{Schematic illustration of the transition from a homogeneous phase to
the bipartite CDW phase.  The hopping is between nearest neighbors, which
corresponds to the neighboring points in the horizontal and vertical
directions.
}
\label{fig: CDW_ordering}
\end{figure}

In order to develop the formalism to determine the Green's functions and
transport in the ordered CDW phase, we need to introduce some notation that will
help clarify how the ordered phase is determined.  It is convenient to 
supplement the lattice site index, which we had been calling $i$, by a 
double index $(i,a)$, where $i$ runs over all of the lattice sites of one
of the sublattices, and the label $a=A$ or $B$ denotes the sublattice
(see Fig.~\ref{fig: CDW_ordering}; we are assuming for simplicity that the
two sublattices have an equal number of lattice sites as they do on the
infinite-dimensional hypercubic lattice or on the infinite-coordination-number
Bethe lattice).  We rewrite the Hamiltonian from Eq.~(\ref{eq: ham_def}) as
\begin{equation}
   \mathcal{\hat{H}}=\sum_{ia}\mathcal{\hat{H}}_{i}^{a}-
   \sum_{ijab}t_{ij}^{ab}\hat{d}_{ia}^{\dag}\hat{d}_{jb},
   \label{eq: ham_def2}
\end{equation}
with the local Hamiltonian satisfying
\begin{equation}
   \mathcal{\hat{H}}_{i}^{a}=U\hat{n}_{id}^{a}\hat{n}_{if}^{a}-
   \mu_{d}^{a}\hat{n}_{id}^{a}-\mu_{f}^{a}\hat{n}_{if}^{a};
\label{eq: ham_loc2}
\end{equation}
in this notation, the bipartite lattice condition is simply that $t_{ij}^{AA}=
t_{ij}^{BB}=0$.
We have introduced different chemical potentials for the two different
sublattices at the moment.   This is convenient for computations, because
it allows us to work with a fixed order parameter, rather than iterating the
DMFT equations to determine the order parameter (which is subject to critical
slowing down near $T_c$).  Of course, the equilibrium solution occurs when
the chemical potential is uniform throughout the system ($\mu^A_d=\mu^B_d$ and
$\mu^A_f=\mu^B_f$).

Our starting point is to find the set of equations satisfied by the lattice
Green's function.  The Green's function is defined to be      
\begin{equation}
G_{ij}^{ab}(\tau)=-\Tr\left [ \mathcal{T}_\tau e^{-\beta\mathcal{\hat H}}
\hat d_{ia}^{}(\tau) \hat d^\dagger_{jb}(0)\right ] / \mathcal{Z},
\label{eq: green_def}
\end{equation}
where $\tau$ is the imaginary time, the time dependence of the destruction 
operator is written in the Heisenberg representation \{$d_{ia}^{}(\tau)=
\exp[\tau\mathcal{\hat H}]d_{ia}^{}\exp[-\tau\mathcal{\hat H}]$\}, 
and $\mathcal{Z}$
is the partition function $\mathcal{Z}=\Tr\exp[-\beta\mathcal{\hat H}]$,
with $\beta=1/T$ the inverse temperature.  The symbol $\mathcal{T}_\tau$ is
the time-ordering operator, which orders the times so that earlier times
appear to the right.  

One way to calculate the Green's function is to use an equation of 
motion technique\cite{zlatic_review}, where the derivative with respect to
imaginary time is taken and a differential equation is found for the
Green's function.  In DMFT, this procedure is carried out for the impurity
problem in a time-dependent field, and the field is adjusted so that the
impurity Green's function is equal to the local lattice Green's function.
In addition, we need to define the self-energy via Dyson's equation in order
to complete the iterative DMFT loop needed to solve the full problem.  Finally,
an analytic continuation from the imaginary axis to the real axis is performed
to calculate dynamical properties.  These techniques are all well known and have
been established in the 
literature\cite{brandt_mielsch1,brandt_mielsch2,zlatic_review,freericks_review},
so we provide just a schematic approach to
the derivation, highlighting some key formulas along the way.

The Dyson equation, which can be thought of as defining the self-energy is
\begin{equation}
\sum_{lc}[(\omega+\mu^{a}_{d})\delta_{ac}\delta_{il}-\Sigma_{il}^{ac}(\omega)
    +t^{ac}_{il}] G_{lj}^{cb}(\omega)
    =\delta_{ij}\delta_{ab},
\label{eq: dyson}
\end{equation}
with $\omega$ the real frequency.
In the case of nearest-neighbor hopping on an infinite-dimensional 
hypercubic lattice, we have that the band structure satisfies
$\epsilon_{\bm k}=-\sum_{j}\exp[i{\bf k}\cdot ({\bf R}_{iA}-{\bf R}_{jB})]
t_{ij}^{AB}=-\lim_{D\rightarrow\infty}
t^{*}\sum\limits_{\alpha=1}^D\cos{k_{\alpha}}/\sqrt{D}$,
where we scaled\cite{metzner_vollhardt} the nearest neighbor hopping 
matrix element by $t=t^*/2\sqrt{D}$ (we will use $t^*=1$ as our energy
unit).  In addition, the self-energy is local\cite{metzner}
\begin{equation}
   \Sigma_{ij}^{ab}(\omega)=\Sigma_{i}^{a}(\omega)\delta_{ij}\delta_{ab},
\label{eq: sigma_local}
\end{equation}
which further simplifies the Dyson equation.  It is simpler to transform from
real space to momentum space to solve the Dyson equation.  But we do not
assume that the Green's function is completely translation invariant, instead,
we assume only that there is translation invariance within each of the
sublattices.  Then the momentum representation of the Dyson equation
in Eq.~(\ref{eq: dyson}) with the local self-energy in 
Eq.~(\ref{eq: sigma_local}) becomes
\begin{equation}
   {G}_{\bm k}(\omega)=\left[{z}(\omega)-{t}_{\bm k}\right]^{-1},
   \label{eq: matrix_dyson}
\end{equation}
where $z(\omega)$ and the hopping term are represented by $2\times 2$ matrices
\begin{align}
   {z}(\omega)&=\left ( \begin{array}{cccc}
      \omega+\mu^{A}_{d}-\Sigma^{A}(\omega) & 0  \\
      0 & \omega+\mu^{B}_{d}-\Sigma^{B}(\omega) \\
   \end{array}\right ),
   \nonumber\\
   {t}_{\bm k}&=\left (\begin{array}{cccc}
      0 & \epsilon_{\bm k}  \\
      \epsilon_{\bm k} & 0 \\
   \end{array}\right ).
\label{eq: twobytwo}
\end{align}
Substituting Eq.~(\ref{eq: twobytwo}) into Eq.~(\ref{eq: matrix_dyson})  and 
taking the matrix inverse yields the following formulas for the 
momentum-dependent Green's functions on the lattice
\begin{align}
   G_{\bm k}^{AA}(\omega)&=
   \dfrac{\omega+\mu^{B}_{d}-\Sigma^{B}(\omega)}{\bar{Z}^{2}(\omega)-\epsilon^2_{\bm k}},
   \label{eq: gaa}\\
   G_{\bm k}^{BB}(\omega)&=
   \dfrac{\omega+\mu^{A}_{d}-\Sigma^{A}(\omega)}{\bar{Z}^{2}(\omega)-\epsilon^2_{\bm k}},
   \label{eq: gbb}\\
   G_{\bm k}^{AB}(\omega)&=G_{k}^{BA}(\omega)=
   \dfrac{\epsilon_{\bm k}}{\bar{Z}^{2}(\omega)-\epsilon^2_{\bm k}}
   \label{eq: gab}
\end{align}
with $\bar Z$ defined by
\begin{equation}
   \bar{Z}(\omega)=\sqrt{[\omega+\mu^{A}_{d}-\Sigma^{A}(\omega)][\omega+\mu^{B}_{d}-\Sigma^{B}(\omega)]},
\label{eq: barz_def}
\end{equation}
which agree with those of Brandt and Mielsch\cite{brandt_mielsch2} even though
our notation is somewhat different from theirs.
The local Green's functions on each sublattice then satisfy
\begin{equation}
G^{aa}(\omega)=\dfrac{\omega+\mu^{b}_{d}-\Sigma^{b}(\omega)}{\bar{Z}(\omega)}
F_z(\omega),
\label{eq: g_loc_sublattice}
\end{equation}
where the $a$ sublattice is different from the $b$ sublattice and
$F_z(\omega)$ is the Hilbert transform
\begin{equation}
F_z(\omega)=\int d\epsilon \rho(\epsilon)\frac{1}{\bar{Z}(\omega)-\epsilon}.
\end{equation}
The function
$\rho(\epsilon)$ is the noninteracting density of states, which
is $\rho(\epsilon)=\exp(-\epsilon^2/t^{*2})/t^*\sqrt{\pi}$ for the
infinite-dimensional hypercubic lattice (as discussed above, we take $t^*=1$).

In the DMFT solution, we need to map the lattice problem onto a local
(impurity) problem in a time-dependent field that is adjusted to make the
impurity Green's function equal to the local Green's function of the
lattice.  Here, we have two different local Green's functions, one
on the A sublattice and one on the B sublattice; hence we will need two
time dependent fields and two impurity problems to solve in order to
complete the DMFT mapping.  We call the dynamical mean fields $\lambda^a(\omega)$
for each sublattice.  Then the solution of the impurity problem is 
straightforward and is summarized by the following set of equations
\begin{align}
G^{a}_0(\omega)&=\left [ G^{aa}(\omega)^{-1}+\Sigma^a(\omega)\right ]^{-1}
\label{eq: g_loc_0}
\\
&=\frac{1}{\omega+\mu^a_d-\lambda^a(\omega)},
\label{eq: lambda}
\\
G^{aa}(\omega)&=\frac{(1-n_f^a)}{\omega+\mu^a_d-\lambda^a(\omega)}\nonumber\\
&+ \frac{n_f^a}{\omega+\mu^a_d-U-\lambda^a(\omega)},
\label{eq: g_loc_imp}
\\
\Sigma^a(\omega)&=\omega+\mu_d^a-\lambda^a(\omega)-G^{aa}(\omega)^{-1},
\label{eq: self_energy_dyson}
\end{align}
where we must solve these equations for each of the sublattices $a=A$ and $a=B$.

The DMFT algorithm for a fixed value of the order parameter starts by
choosing $n_f^A$ and $n_f^B$ such that $n_f^A+n_f^B$ is fixed to the total
$f$-electron filling (the order parameter is $\Delta n_f=n_f^A-n_f^B$), 
and choosing $\mu_d^A=\mu_d^B$.  With those fixed
quantities, we propose a guess for the self-energy on each sublattice, and
then compute the local Green's function on the real axis from 
Eqs.~(\ref{eq: barz_def}) and (\ref{eq: g_loc_sublattice}).  Then we
extract the dynamical mean field on each sublattice from Eqs.~(\ref{eq: g_loc_0})
and (\ref{eq: lambda}), then find the local Green's function for the
impurity from Eq.~(\ref{eq: g_loc_imp}) and the new self-energy from
Eq.~(\ref{eq: self_energy_dyson}).  This loop is repeated until the
Green's functions converge.  Then one can calculate the filling of
the $d$-electrons and adjust them until they match the target filling.
But this procedure is not yet complete, because we need to determine the
correct equilibrium order parameter $n_f^A-n_f^B$ at the given temperature.
To find this, it is actually more convenient to perform the calculations
precisely as described above, but on the imaginary frequency axis, where
$\omega$ is replaced by $i\omega_n=i\pi T(2n+1)$ the fermionic Matsubara
frequencies. Then we calculate the chemical potential for the $f$-electrons
on each sublattice via
\begin{equation}
\mu^a_f=-\frac{U}{2}-T\ln \frac{1-n_f^a}{n_f^a}-T\sum_n\ln [1-UG_0^a(i\omega_n)],
\label{eq: chem_pot_imag}
\end{equation}
and adjust the order parameter until the two chemical potentials are equal,
which is required for the equilibrium solution. Then, when we calculate the
Green's functions on the real axis, the chemical potentials and fillings are
all already known, so they do not need to be adjusted during the calculation.

This algorithm is much more efficient than an algorithm that starts with
a fixed chemical potential for the $f$-electrons and iterates to determine
$n_f^a$ on the imaginary frequency axis.  This is because the latter suffers from
critical slowing down, and becomes quite inefficient near the critical
temperature, whereas the calculations with the fixed order parameter converge
quite rapidly regardless of how close one is to the critical point.

When the DOS is calculated for each sublattice in the ordered phase,
one finds interesting temperature dependence of the subgap states as 
a function of $T$.  It is illustrative to discuss these evolutions in terms
of the moments of the local interacting DOS.  It is well known, that in the
homogeneous phase, the integral of $A(\omega)=-\Img G(\omega)/\pi$ is equal to
1. But there are also exact results known for higher moments as 
well\cite{white_moment,freericks_moment}.  In particular, because the moments
are derived from operator identities, they continue to hold whether in the 
ordered phase or not.  So we learn the following identities immediately:
\begin{align}
\int d\omega A^a(\omega)&=1;\label{eq: zeroth_moment}\\
\int d\omega \omega A^a(\omega)&=-\mu_d^a+Un_f^a;\label{eq: first_moment}\\
\int d\omega \omega^2A^a(\omega)&=\frac{1}{2}+\mu_d^{a2}-2U\mu_d^a n_f^a+
U^2n_f^a.
\label{eq: second_moment}
\end{align}
We have checked these moments versus our numerical calculations of the Green's
functions on the real axis and they all agree
to high accuracy for all temperatures that we consider.  Note that at half 
filling, we have $\mu_d^a=U/2$, so the first moment vanishes in the
homogeneous phase.  As the system orders, the first moment on one sublattice
becomes negative, and the first moment on the other sublattice becomes
positive, which indicates that the quantum states are shifting
in response to the ordering.  In particular, this redistribution of states 
causes the average kinetic energy to evolve more strongly with temperature
in the ordered phase, but its evolution is anomalous, and cannot be predicted
by any simple reasoning about how the states evolve (see below). The evolution
of the average kinetic energy plays an important role in the total spectral
weight for the optical conductivity.

\begin{figure}[tb]
\includegraphics[width=0.4\textwidth]{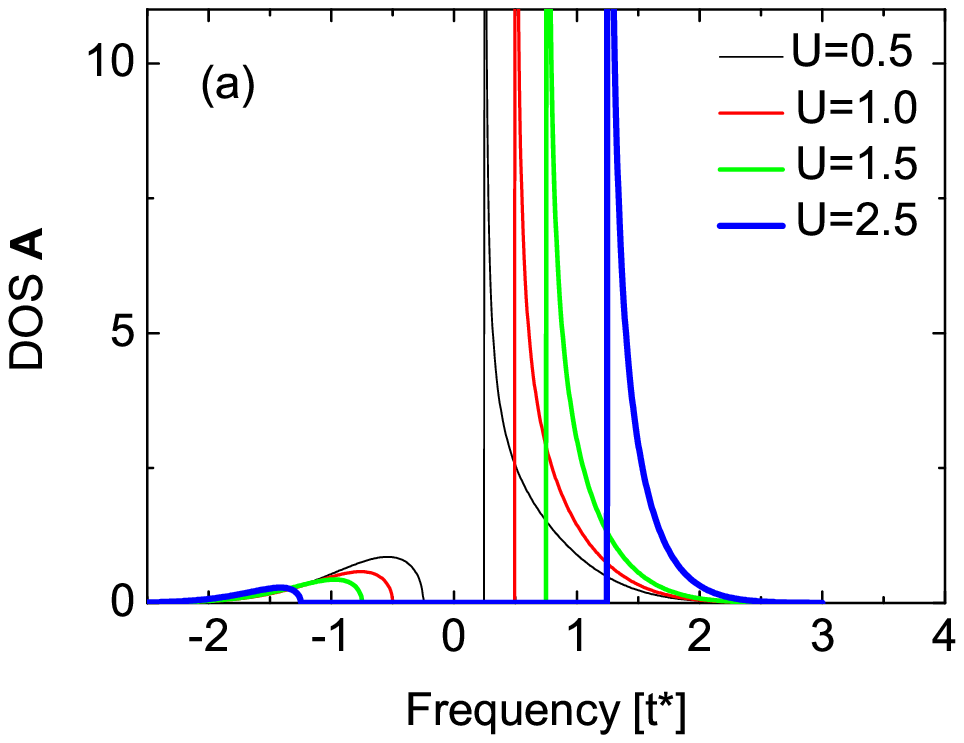}\\
\includegraphics[width=0.4\textwidth]{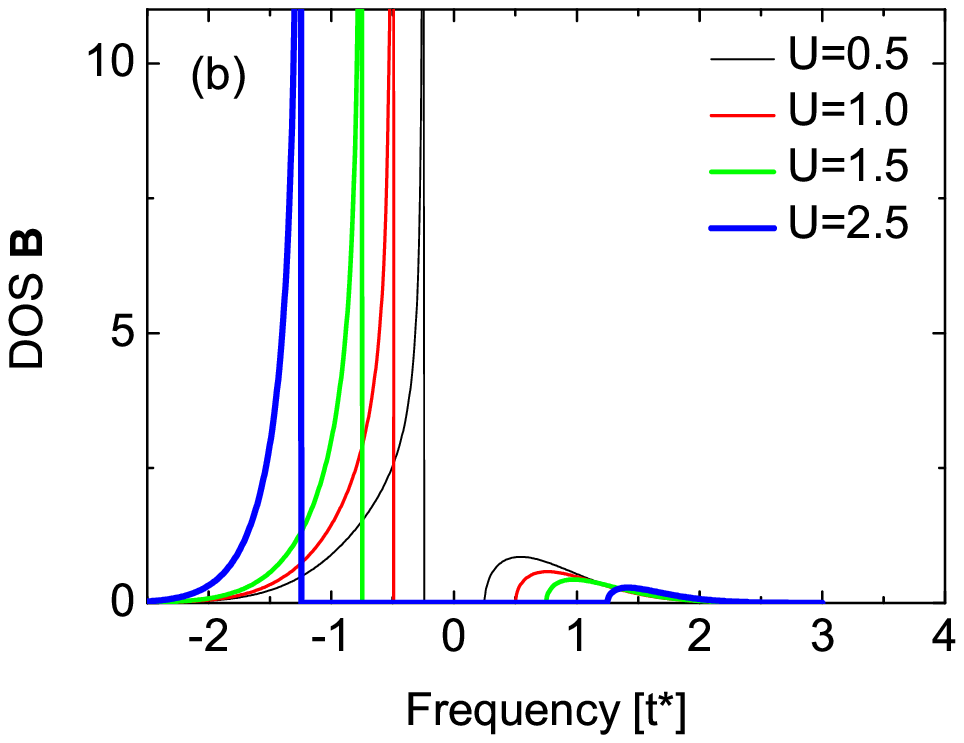}\\
\caption{(Color online) DOS at $T=0$ for the CDW-ordered phase on a hypercubic lattice.
Panel (a) is for the $A$ sublattice and panel (b) is for the $B$ sublattice.
Four cases are plotted: $U=0.5$ which is a strongly correlated metal;
$U=1$, where a dip develops in the normal-state DOS at the chemical
potential;
$U=1.5$, which is a near-critical Mott insulator; and $U=2.5$, which is
a moderate-size-gap Mott insulator.  The $T=0$ gap in the DOS is
always equal to $U$ in the ordered CDW phase.
\label{fig: DOS}}
\end{figure}

At $T=0$, the order parameter goes to 1, so there is one sublattice (let us say
the $A$ sublattice) which has all the $f$-electrons.  Hence $n_f^A=1$ and
$n_f^B=0$.  In this case, the analysis for the Green's function simplifies.
In particular, only one term in Eq.~(\ref{eq: g_loc_imp}) survives on each
sublattice and we immediately find $\Sigma^A=U$ and $\Sigma^B=0$. Plugging
these results into the remaining formulas for the DMFT algorithm then yields
an analytic formula for the ordered phase DOS
\begin{align}
A_{A,B}(\omega)&=-\frac{1}{\pi}\Img G^{AA,BB}(\omega)\nonumber\\
&=\Real\left [ \sqrt{\frac{\omega\pm\frac{U}{
2}}{\omega\mp\frac{U}{2}}}\right ]
\rho\left(\sqrt{\omega^2-\frac{U^2}{4}}\right),
\label{eq: dos_t=0}
\end{align}
where the top sign is for the $A$ sublattice (with a divergence of the
DOS at $\omega=U/2$) and the bottom sign is for the $B$ sublattice
(with a divergence at $\omega=-U/2$); the formula is restricted to half filling
where $\mu_d^A=\mu_d^B=U/2$. Note that the two DOS on each sublattice
are mirror images of each other and that each sublattice has weight for 
positive and negative frequency, but the band that does not have the
singularity (lower band for sublattice $A$ and upper band for sublattice
$B$) has shrinking spectral weight as $U$ becomes large, because the mobile
electrons avoid the sites with the localized electrons for large $U$. Note 
further, that unlike the Mott insulator, where the DOS vanishes only at the
chemical potential on a hypercubic lattice, a real gap develops here of
magnitude $U$ at $T=0$.
In Fig.~\ref{fig: DOS}, we show the DOS at zero temperature for
four values of $U$. Panel (a) plots the DOS on the $A$ sublattice and
panel (b) plots the DOS on the $B$ sublattice.
One can see that the shape of the DOS is qualitatively similar for all
cases, but the size of the gap grows with $U$.

\begin{figure}[tb]
\includegraphics[width=0.4\textwidth]{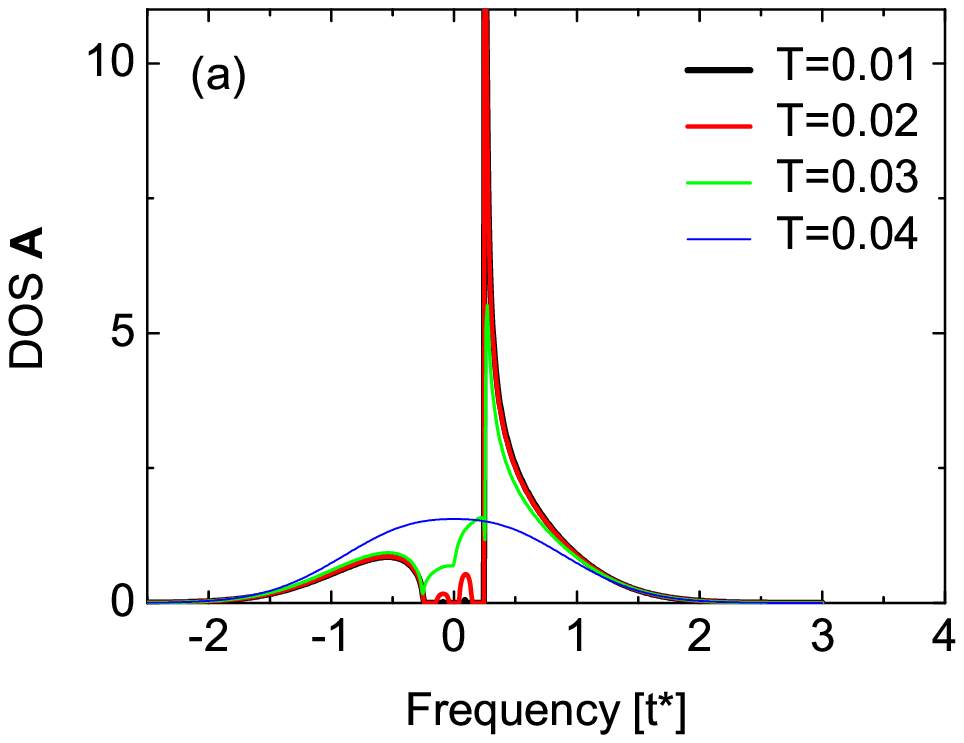}\\
\includegraphics[width=0.4\textwidth]{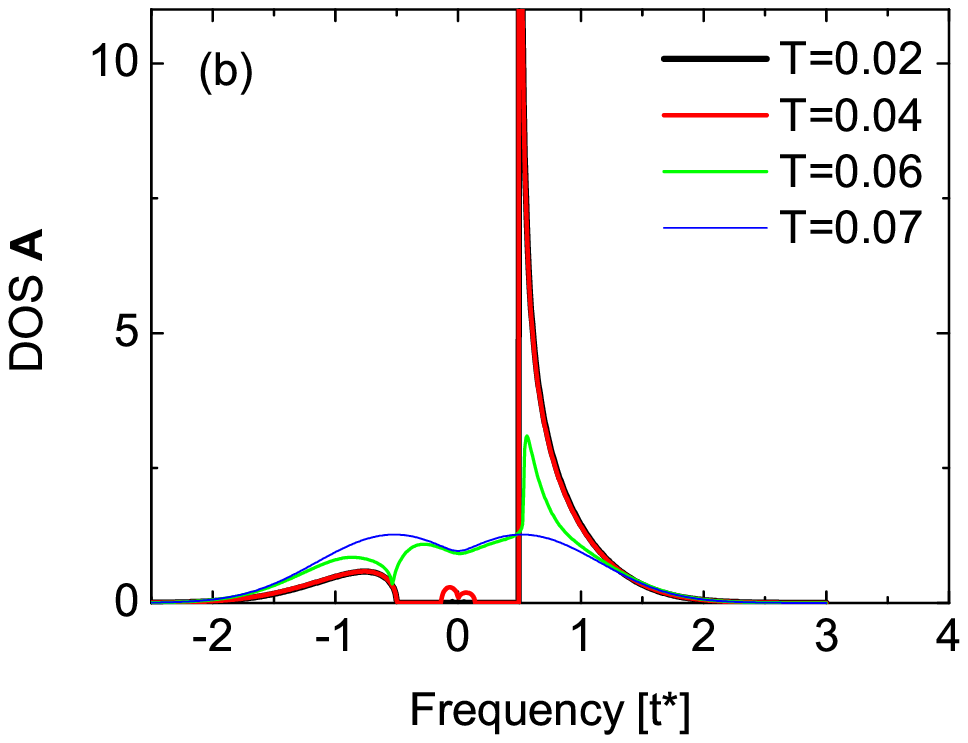}
\caption{(Color online) DOS on the A sublattice for various $T$ values in the CDW-ordered 
phase on a hypercubic lattice with (a) $U=0.5$ and (b) $U=1$.
The DOS on the B sublattice is a mirror reflection of these results about the
plane $\omega=0$. 
\label{fig: u=0.5_dos}}
\end{figure}

What is more interesting is to examine the temperature evolution of the DOS
in these different cases.  Indeed, the system develops substantial subgap
DOS that is thermally excited within the ordered phase (the order parameter
is determined by the difference in localized electron filling on the two
sublattices).  In Fig.~\ref{fig: u=0.5_dos}~(a), we plot the DOS for the
strongly correlated metal at $U=0.5$.  The fill in of the subgap states is quite
rapid with $T$ as we increase up to $T_c=0.0336$. Similar behavior is also
observed for $U=1$ with $T_c=0.0615$ which has a dip in the DOS in the normal 
state [Fig.~\ref{fig: u=0.5_dos}~(b)].

\begin{figure}[htb]
\includegraphics[width=0.4\textwidth]{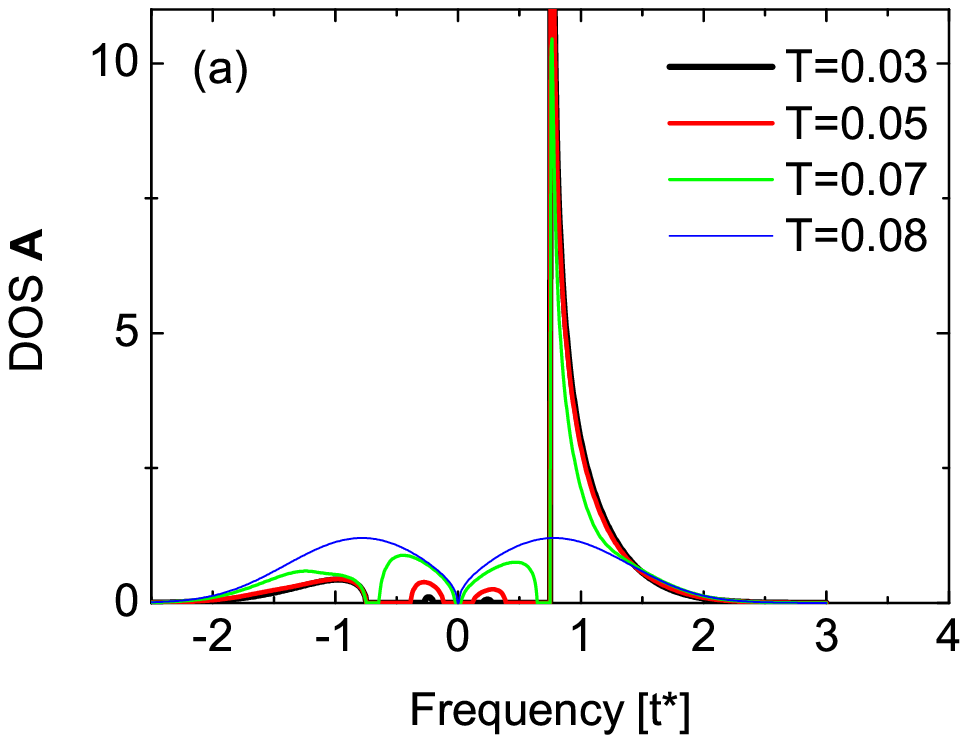}\\
\includegraphics[width=0.4\textwidth]{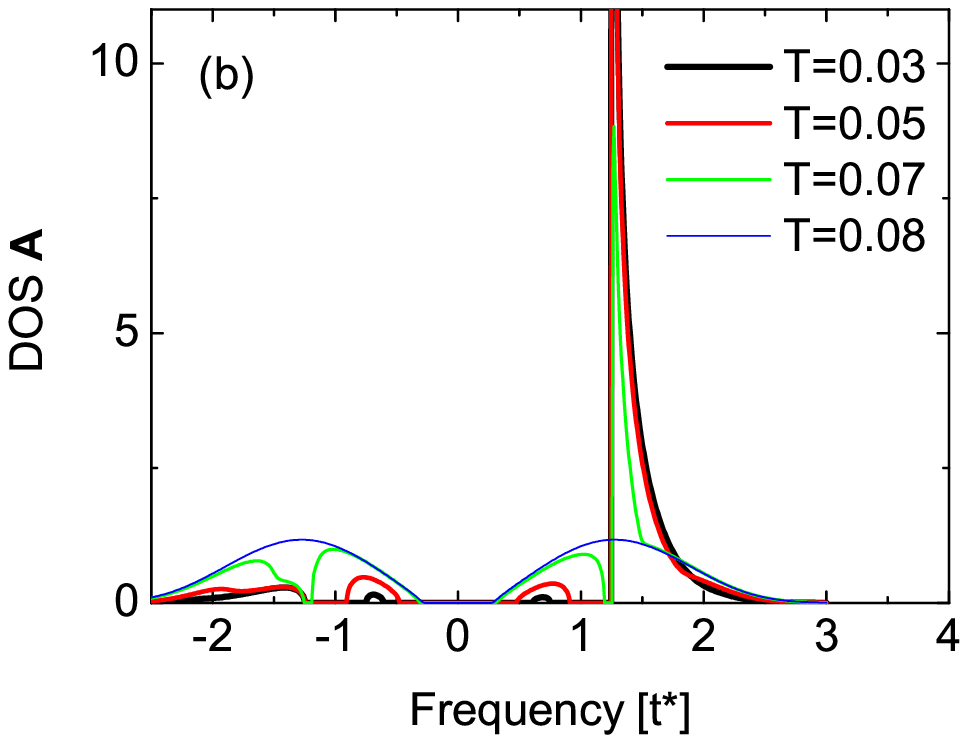}
\caption{(Color online) DOS on the A sublattice for various $T$ values in the CDW-ordered 
phase on a hypercubic lattice with (a) $U=1.5$ and (b) $U=2.5$.
\label{fig: u=1.5_dos}}
\end{figure}

The Mott insulating phases also illustrate interesting behavior.  In
particular, the subgap states develop primarily within the upper and
lower Hubbard bands (although on the hypercubic lattice, the Mott insulator
has only a pseudogap with the DOS strictly vanishing only at $\omega=0$).  
We illustrate this behavior in Figs.~\ref{fig: u=1.5_dos}~(a) and (b).
The transition temperatures are $T_c=0.0747$ for
$U=1.5$ and $T_c=0.0724$ for $U=2.5$.  Note how the subgap DOS develop closer
to the Mott band edge than they do to the CDW band edge, which implies they
should have an effect on the transport at low $T$.

In all cases, the DOS satisfy the three sum rules for the first three moments
to essentially machine accuracy---our actual accuracy is determined by the
step size we use for the real frequency axis in calculating the DOS and
then integrating it over all frequency to obtain the numerical moments.

Now we develop the formalism for transport in the CDW phase.
The linear response optical conductivity is determined (via the Kubo-Greenwood
formula\cite{kubo,greenwood}) by the
imaginary part of the analytic continuation of the
current-current correlation function to the real axis,
\begin{equation}
   \sigma(\omega)=\dfrac{1}{\omega}\Img{\Pi_{jj}(\omega)},
\end{equation}
with the number current operators defined by
\begin{align}
   {\bm{\hat j}} &= i\sum_{ijab} t_{ij}^{ab} (\bm R_{ia}-\bm R_{jb})
             \hat{d}_{ia}^{\dag}\hat{d}_{jb},\\
   j_{\alpha}&=\sum\limits_{ab\bm k}
   \dfrac{\partial\epsilon_{\bm k}^{ab}}{\partial
   k_{\alpha}}d_{a}^{\dag}(\bm k)d_{b}(\bm k).
   \label{j}
\end{align}
The procedure to determine the current-current correlation function is
a standard one so we only sketch the derivation briefly.  
We start from the imaginary time formula for the current-current correlation
function
\begin{equation}
   \Pi_{jj}(\tau-\tau^\prime)=\left\langle T_{\tau}j(\tau)j(\tau^\prime)
\right\rangle,
   \label{gjj}
\end{equation}
where the angle brackets denote a trace over all states weighted by the
statistical operator (density matrix) at the given temperature
and the current operators are represented in the Heisenberg 
representation with respect to the equilibrium Hamiltonian (because
this is a linear-response calculation).
We then perform a Fourier transformation to go from imaginary time
to Matsubara frequencies, and then perform an analytic continuation from
the imaginary frequency axis to the real frequency axis.

\begin{figure}[htb]
   \includegraphics[height=.08\textheight]{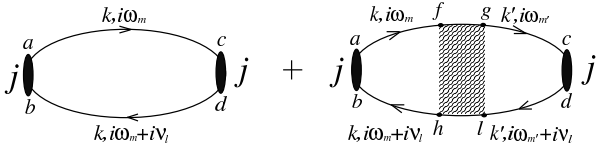}
   \caption{\label{fig: response} Bethe-Salpeter equation for the 
generalized polarization.}
\end{figure}

The Fourier transform of the current-current
correlation function defined in Eq.~(\ref{gjj}) can be represented as
a summation over Matsubara frequencies
\begin{equation}
   \Pi_{jj}(i\nu_l)=T\sum\limits_{m}\Pi_{m,m+l}
   \label{Pi}
\end{equation}
where we introduced the shorthand notation
$\Pi_{m,m+l}=\Pi(i\omega_{m},i\omega_{m}+i\nu_l)$ for the dependence on the
fermionic 
$i\omega_m=i\pi T(2m+1)$ and bosonic $i\nu_l=i2\pi T l$ Matsubara 
frequencies ($m$ and $l$ are integers).
In the CDW phase, the graphic depiction of the Bethe-Salpeter equation
for the generalized polarization $\Pi_{m,m+\nu}$ is plotted in 
Fig.~\ref{fig: response}
where the solid oval depicts the current operator using the same sublattice
indices as we have used before (the current operator connects the two
sublattices), the solid lines are Green's functions, and the cross hatched object
is the total (reducible) charge vertex.  
The current operator vertex contains the factor 
$\partial\epsilon_{\bm k}/\partial k_\alpha$ which is an odd function of 
the wavevector.  Since the band structure $\epsilon_{\bm k}$ and the Green's 
functions are even functions of the wavevector, any summation over momentum that 
contains one current vertex and any number of Green's functions will vanish. 
Now, in infinite dimensions,
the irreducible charge vertex  (which enters 
the Bethe-Salpeter equation for the total charge vertex) is local and hence momentum independent, so
the second term in Fig.~\ref{fig: response} vanishes, just like it did
in the homogeneous phase\cite{khurana}.
We thereby conclude that the optical
conductivity is constructed only by the bare bubble in Fig.~\ref{fig: response}. 
\begin{figure}[htb]
   $\Large{\Pi_{m,m+l}=\sum\limits_{abcd}}$
   \raisebox{-.45\height}{\includegraphics[height=.07\textheight]{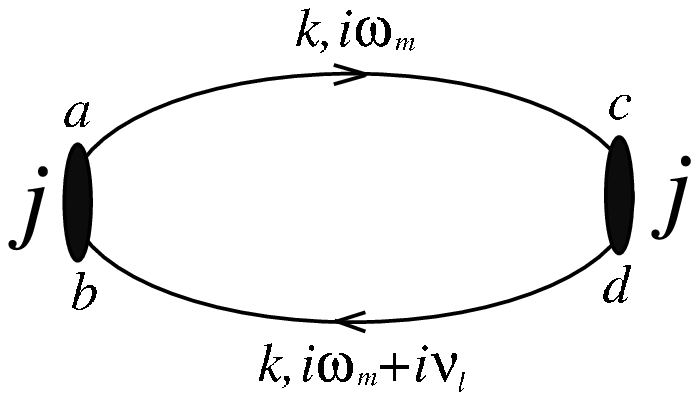}}
   \\
   =\raisebox{-.45\height}{\includegraphics[height=.07\textheight]{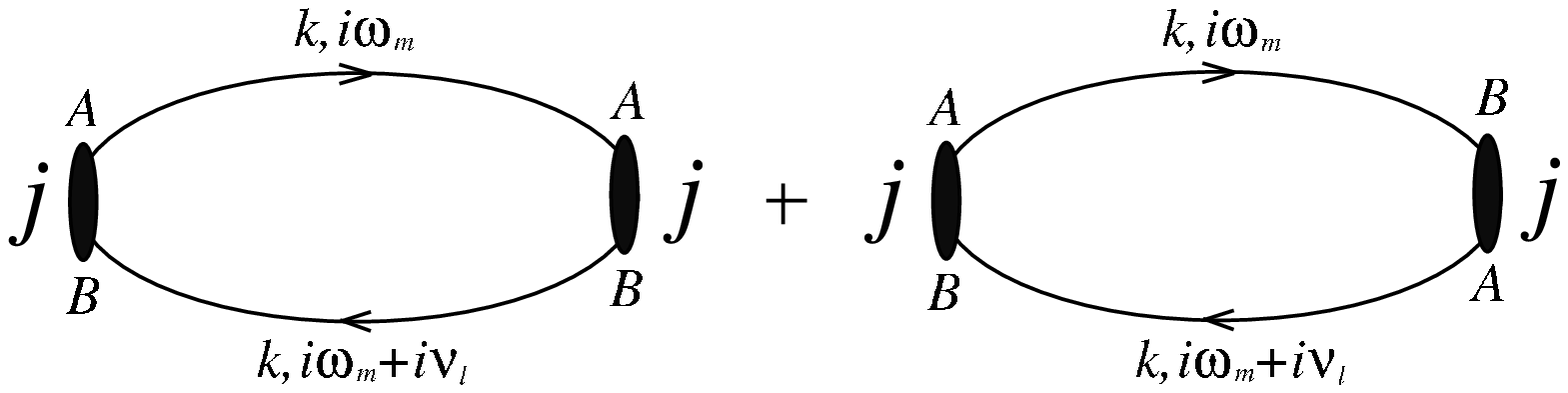}}
   \\
   +\raisebox{-.45\height}{\includegraphics[height=.07\textheight]{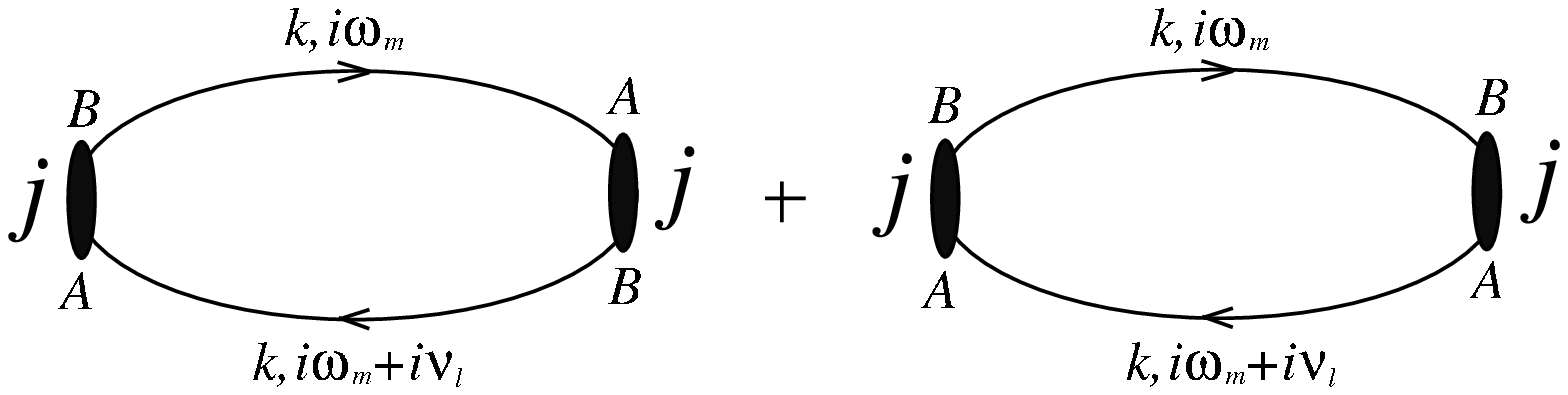}}
\caption{Individual terms for the bare polarization in the ordered phase.
\label{fig: polarization}}
\end{figure}
Then, the full expression for
the generalized polarization $\Pi_{m,m+l}$ is depicted in 
Fig.~\ref{fig: polarization} and simplifies to
\begin{align}\label{sum}
   \Pi_{m,m+l}
   &=\frac{1}{N}\sum_{\bm k} j_{\bm k}^{2}(G_{\bm km}^{AA}G_{\bm km+l}^{BB}+
   G_{\bm km}^{AB}G_{\bm km+l}^{AB}\nonumber\\
&+G_{\bm km}^{BA}G_{\bm km+l}^{BA}+
   G_{\bm km}^{BB}G_{\bm km+l}^{AA}),
\end{align}
where $j_{\bm k}=-\lim_{D\rightarrow\infty}(t^*/\sqrt{D})\sum_{r=1}^D \sin k_r$ 
and solid lines denote the momentum-dependent lattice Green's functions 
$G_{\bm km}^{ab}$ [see Eqs.~(\ref{eq: gaa}--\ref{eq: gab})].
After substituting in the expressions for the Green's functions, 
the individual contributions to $\Pi_{m,m+l}$ become
\begin{align}
   &\dfrac{1}{N}\sum\limits_{\bm k} j_{\bm k}^{2} G_{\bm km}^{AA}G_{\bm km+l}^{BB}
   =\dfrac{1}{2}(i\omega_m+\mu_{d}^{B}-\Sigma_{m}^B)
\nonumber\\
\times&(i\omega_m+i\nu_l+\mu_{d}^{A}-\Sigma_{m+l}^A)
   \dfrac{\dfrac{F_z(i\omega_{m+l})}{\bar{Z}(i\omega_{m+l})}
-\dfrac{F_z(i\omega_m)}{\bar{Z}(i\omega_m)}}{\bar{Z}^{2}(i\omega_m)-
\bar{Z}^{2}(i\omega_{m+l})},
   \nonumber
\end{align}
\begin{align}
   &\dfrac{1}{N}\sum\limits_{\bm k} j_{\bm k}^{2} G_{\bm km}^{BB}G_{\bm km+l}^{AA}
   =\dfrac{1}{2}(i\omega_m+\mu_{d}^{A}-\Sigma_{m}^A)\nonumber\\
\times&(i\omega_m+i\nu_l+\mu_{d}^{B}-\Sigma_{m+l}^B)
   \dfrac{\dfrac{F_z(i\omega_{m+l})}{\bar{Z}(i\omega_{m+l})}
-\dfrac{F_z(i\omega_m)}{\bar{Z}(i\omega_m)}}{\bar{Z}^2(i\omega_m)
-\bar{Z}^{2}(i\omega_{m+l})},
   \nonumber
\end{align}
\begin{align}
   &\dfrac{1}{N}\sum\limits_{\bm k} j_{\bm k}^{2} G_{\bm km}^{AB}G_{\bm km+l}^{AB}
   =\dfrac{1}{N}\sum\limits_{\bm k} j_{\bm k}^{2} G_{\bm km}^{BA}G_{\bm k m+l}^{BA}
\nonumber\\
   {}=&\dfrac{1}{2}\dfrac{\bar{Z}(i\omega_{m+l})
F_z(i\omega_{m+l})-\bar{Z}(i\omega_m)F_z(i\omega_m)}
{\bar{Z}^{2}(i\omega_m)-\bar{Z}^{2}(i\omega_{m+l})}.
\end{align}
Hence, the full expression for $\Pi_{m,m+l}$ is 
\begin{align}\label{b1gPi}
   \Pi_{m,m+l}&=\dfrac{1}{2}\Biggl\{
   \frac{\dfrac{F_z(i\omega_{m+l})}{\bar{Z}(i\omega_{m+l})}
-\dfrac{F_z(i\omega_{m})}{\bar{Z}(i\omega_{m})}}{\bar{Z}^2(i\omega_{m})
-\bar{Z}^2(i\omega_{m+l})}\nonumber\\
   &\times\biggl[(i\omega_m+\mu_{d}^{B}-\Sigma_{m}^B)
(i\omega_m+i\nu_l+\mu_{d}^{A}-\Sigma_{m+l}^A)\nonumber\\
   &+(i\omega_m+\mu_{d}^{A}-\Sigma_{m}^A)(i\omega_m+i\nu_l+\mu_{d}^{B}
-\Sigma_{m+l}^B)\biggr]\nonumber\\
   &+2\dfrac{\bar{Z}(i\omega_{m+l})F_z(i\omega_{m+l})
-\bar{Z}(i\omega_m)F_z(i\omega_m)}{\bar{Z}^2(i\omega_m)-\bar{Z}^2(i\omega_{m+l})}\Biggr\}.
\end{align}
Then, the expression for the current-current Green's function is obtained by
substituting Eq.~(\ref{b1gPi}) into  Eq.~(\ref{Pi}) and analytically continuing
the summation over Matsubara frequencies into contour integrations
\begin{align}
   \Pi_{jj}(i\nu_l)&=\frac{1}{2\pi i}
   \int\limits_{-\infty}^{+\infty}d\tilde\omega f(\tilde\omega)\nonumber\\
&\times\Big[\Pi(\tilde\omega-i0^{+},\tilde\omega+i\nu_l)
-\Pi(\tilde\omega+i0^{+},\tilde\omega+i\nu_l)
   \nonumber\\
   &+\Pi(\tilde\omega-i\nu_l,\tilde\omega-i0^{+})
-\Pi(\tilde\omega-i\nu_l,\tilde\omega+i0^{+})\Big].
   \label{h}
\end{align}
Here we have $f(\tilde\omega)=1/[1+\exp(\beta\tilde\omega)]$ is the fermi
distribution function. The final step is to analytically continue from the
bosonic Matsubara frequencies to the real axis
($i\nu\to\omega\pm i0^{+}$).  This produces our final result
\begin{align}
   \Pi_{jj}(\omega)&=
\dfrac{2}{(2\pi i)^{2}}\int\limits_{-\infty}^{+\infty}d\tilde\omega
   \left[f(\tilde\omega)-f(\tilde\omega+\omega)\right]
   \nonumber\\
   &\times\Real\{\Pi(\tilde\omega-i0^{+},\tilde\omega+\omega+i0^{+})
\nonumber\\
&-\Pi(\tilde\omega-i0^{+},\tilde\omega+\omega-i0^{+})\}.
   \label{hh}
\end{align} 
To make Eq.~(\ref{hh}) concrete, we substitute in the analytic continuation of
Eq.~(\ref{b1gPi}) to find the final expression for the optical
conductivity (we set $e^2=1$):
\begin{align}
   \sigma(\omega)&=\dfrac{1}{4\pi^{2}}\int\limits_{-\infty}^{+\infty}d\tilde\omega
   \dfrac{\left[f(\tilde\omega)-f(\tilde\omega+\omega)\right]}{\omega}
   \nonumber \\
   &\times\Real\Biggl\{\dfrac{\dfrac{F^{*}_z(\tilde\omega+\omega)}{\bar{Z}^{*}(\tilde\omega+\omega)}-\dfrac{F_z(\tilde\omega)}{\bar{Z}(\tilde\omega)}}
   {\bar{Z}^{2}(\tilde\omega)-[\bar{Z}^{*}(\tilde\omega+\omega)]^{2}}
\nonumber\\
&\times
   \biggl([\tilde\omega+\mu_{d}^{B}-\Sigma^{B}(\tilde\omega)]
[\tilde\omega+\omega+\mu_{d}^{A}-\Sigma^{A*}(\tilde\omega+\omega)]
   \nonumber \\
   &+[\tilde\omega+\mu_{d}^{A}-\Sigma^{A}(\tilde\omega)]
[\tilde\omega+\omega+\mu_{d}^{B}-\Sigma^{B*}(\tilde\omega+\omega)]\biggl)
\nonumber\\
   &+2\dfrac{\bar{Z}^{*}(\tilde\omega+\omega)F_z^{*}(\tilde\omega+\omega)-
\bar{Z}(\tilde\omega)F_z(\tilde\omega)}
   {\bar{Z}^{2}(\tilde\omega)-[\bar{Z}^{*}(\tilde\omega+\omega)]^{2}}
   \nonumber\\
   &-\dfrac{\dfrac{F_z(\tilde\omega+\omega)}{\bar{Z}(\tilde\omega+\omega)}
   -\dfrac{F_z(\tilde\omega)}{\bar{Z}(\tilde\omega)}}{\bar{Z}^{2}(\tilde\omega)
-\bar{Z}^{2}(\tilde\omega+\omega)}\nonumber\\
&\times
   \biggl([\tilde\omega+\mu_{d}^{B}-\Sigma^{B}(\tilde\omega)][\tilde\omega+\omega+\mu_{d}^{A}-\Sigma^{A}(\tilde\omega+\omega)]
   \nonumber \\
   &+[\tilde\omega+\mu_{d}^{A}-\Sigma^{A}(\tilde\omega)][\tilde\omega+\omega+\mu_{d}^{B}-\Sigma^{B}(\tilde\omega+\omega)]\biggr)\nonumber\\
   &-2\dfrac{\bar{Z}(\tilde\omega+\omega)F_z(\tilde\omega+\omega)
-\bar{Z}(\tilde\omega)F_z(\tilde\omega)}
   {\bar{Z}^{2}(\tilde\omega)-\bar{Z}^{2}(\tilde\omega+\omega)}\Biggr\}.
   \label{sigma}
\end{align}

\begin{figure*}[htb]
\includegraphics[width=0.4\textwidth]{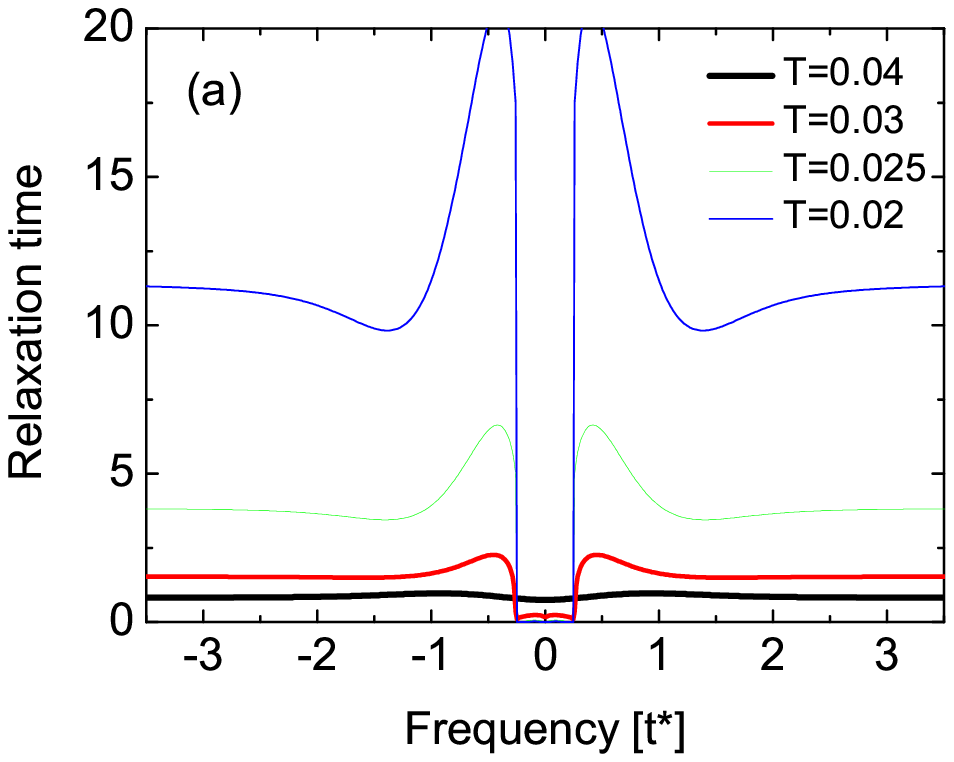}\qquad
\includegraphics[width=0.4\textwidth]{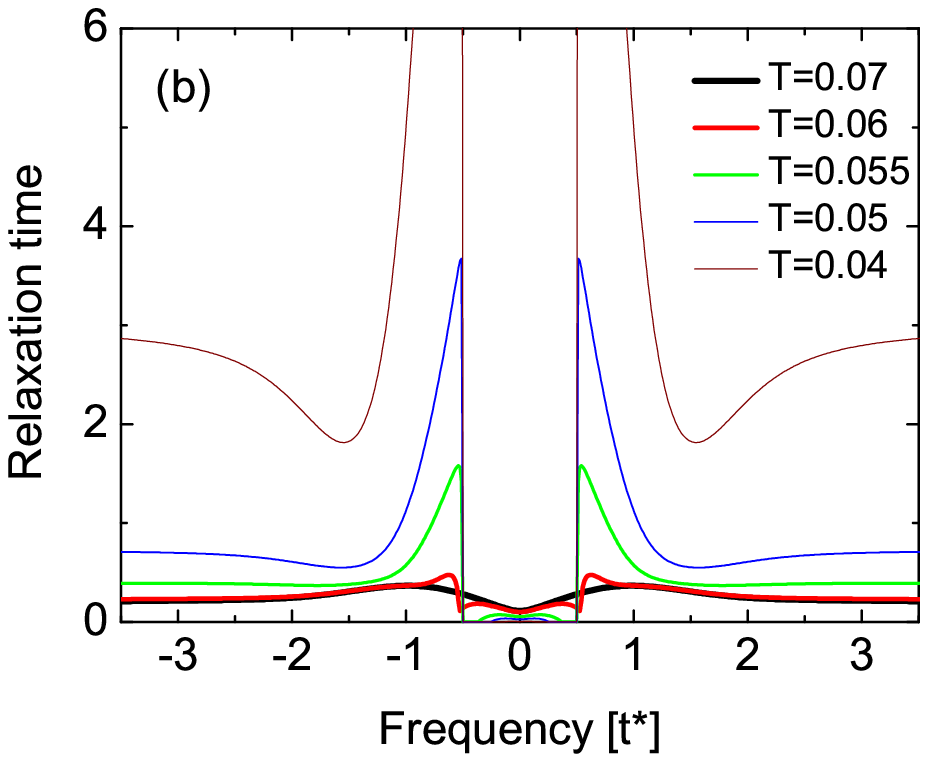}\\
\includegraphics[width=0.4\textwidth]{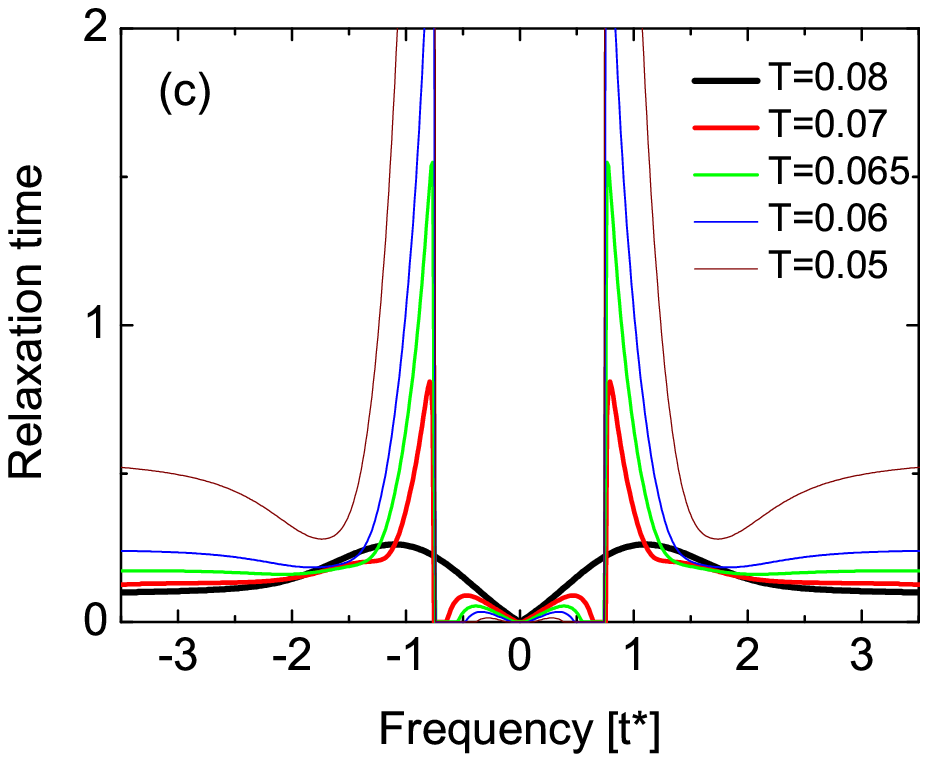}\qquad
\includegraphics[width=0.4\textwidth]{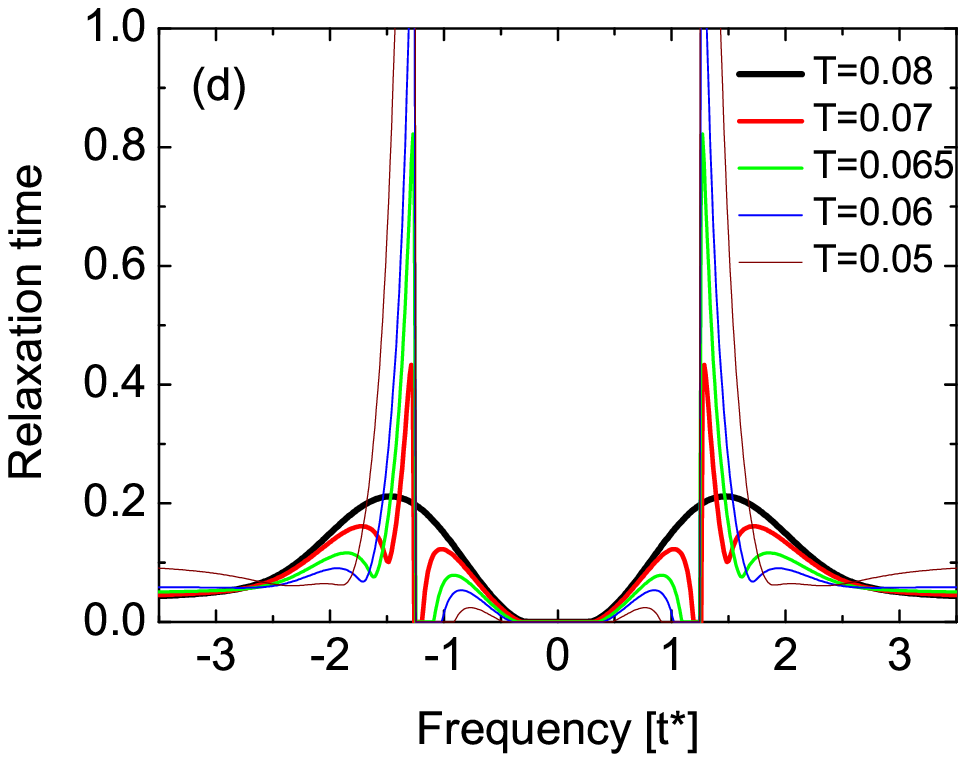}
\caption{(Color online) Exact many-body
relaxation time at various $T$ values for the CDW-ordered phase on a
hypercubic lattice with (a) $U=0.5$, (b) $U=1$, (c) $U=1.5$, and (d) $U=2.5$.
At high $T$ in the Mott-insulator, the relaxation time goes to zero as
$\omega^4$ [this is hard to see in panel (d) because the quartic region 
occurs only for small frequencies and cannot be easily seen on this
linear scale plot]. 
\label{fig:rel_t}}
\end{figure*}

The final formalism we need to develop is for the dc transport properties.
Starting from the expression for the optical conductivity in Eq.~(\ref{sigma}) 
we can calculate the dc conductivity by taking the zero frequency limit:
\begin{equation}
   \sigma_{dc}=\lim\limits_{\omega\to0}\sigma(\omega).
\end{equation}
The algebra is completely straightforward, but requires a careful
use of l'H\^opital's rule for determining some of the limits.  After
some lengthy algebra, we find that
the final expression of the dc conductivity becomes
\begin{equation}\label{dc_cond}
   \sigma_{dc}=2\int\limits_{-\infty}^{+\infty}d\omega
   \left[-\dfrac{df(\omega)}{d\omega}\right] \tau(\omega)
\end{equation}
with the exact many-body relaxation time $\tau(\omega)$ equal to
\begin{widetext}
\begin{align}\label{RT_def}
   \tau(\omega)&=\dfrac{1}{4\pi ^{2}}
   \Biggl\{\dfrac{1}{2}\biggl[
   \dfrac{\Real\left\{ [\omega+\mu_{d}^{A}-
    \Sigma^{A}(\omega)][\omega+\mu_{d}^{B}
   -\Sigma^{B*}(\omega)]\right\}}{|\bar{Z}(\omega)|^2}
   \left\{\dfrac{\Real F_z(\omega)}{\Real \bar{Z}(\omega)}
-\dfrac{\Img F_z(\omega)}{\Img \bar{Z}(\omega)}\right\}
   \nonumber \\
   & -\left\{\dfrac{\Real F_z(\omega)}{\Real
\bar{Z}(\omega)}+\dfrac{\Img F_z(\omega)}{\Img
   \bar{Z}(\omega)}\right\}\biggr]-
2\Real[\bar{Z}(\omega)F_z(\omega)-1]\Biggr\}.
\end{align}
\end{widetext}
For large frequencies the relaxation time approaches the asymptotic value
\begin{equation}
 \tau_{\infty}=\dfrac{1}{4\pi ^{2}}\frac{2}{U^2[n_f^A(1-n_f^A)+n_f^B(1-n_f^B)]};
\end{equation}
this is a well-known anomaly on the infinite-dimensional hypercubic
lattice\cite{demchenko} due to the fact that the DOS never vanishes and at
large frequencies the imaginary part of the self-energy is exponentially 
small, implying very long lifetimes for the excitations. Note that the 
high-frequency limit of $\tau(\omega)$ actually diverges as $T\rightarrow 0$
at half filling.  This trend can be seen to develop in Fig.~\ref{fig:rel_t}, 
although we do not push the calculations too low in temperature due to accuracy 
issues with determining the subgap states.

Starting from Eq.~(\ref{dc_cond}) we can also calculate the thermal
transport.  Since the system is at half-filling, the thermopower
vanishes due to particle-hole symmetry: the relaxation time in 
Eq.~(\ref{RT_def}) is symmetric with respect to sublattice indices and is 
an even function of frequency at half-filling (Fig.~\ref{fig:rel_t}).
The electronic
contribution to the thermal conductivity $\kappa_e$ is nonzero, and can be found
in the standard fashion.  It is expressed in terms of
three different transport coefficients $L_{11}$, $L_{12}=L_{21}$
and $L_{22}$ as follows:~\cite{luttinger}
\begin{equation}
   \kappa_{e}=\dfrac{1}{T}\left[L_{22}-\dfrac{L_{12}L_{21}}{L_{11}}\right].
\end{equation}
In this notation, the dc conductivity satisfies
\begin{equation}
   \sigma_{dc}=L_{11}.
\end{equation}
The other transport coefficients can be calculated from the
Jonson-Mahan theorem\cite{JMT1,JMT2} which says that there is a
simple relation between these different transport coefficients,
namely that they reproduce the so-called Mott-Thellung noninteracting
form\cite{CT},
\begin{equation}
   L_{ij}=\int\limits_{-\infty}^{+\infty}d\omega\left[-\dfrac{d
   f(\omega)}{d\omega}\right]\tau(\omega)\omega^{i+j-2},
\end{equation}
where $\tau(\omega)$ is the exact many-body relaxation time defined 
in Eq.~(\ref{RT_def}) and plotted in Fig.~\ref{fig:rel_t}.

\section{Numerical Results}

\begin{figure}[htb]
 \includegraphics[width=0.4\textwidth]{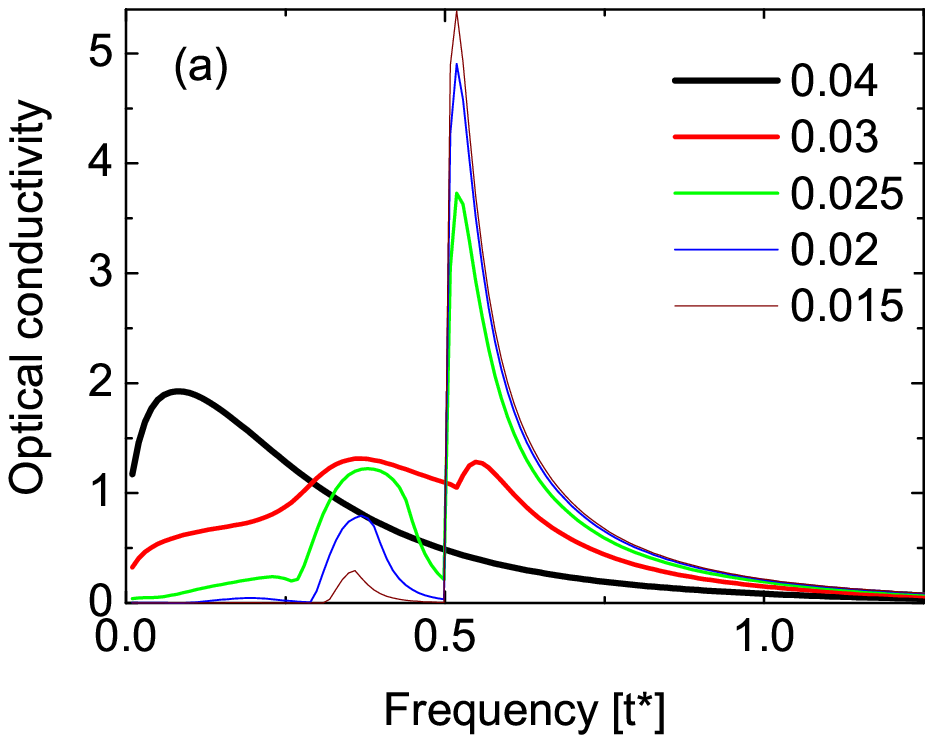}\\
 \includegraphics[width=0.4\textwidth]{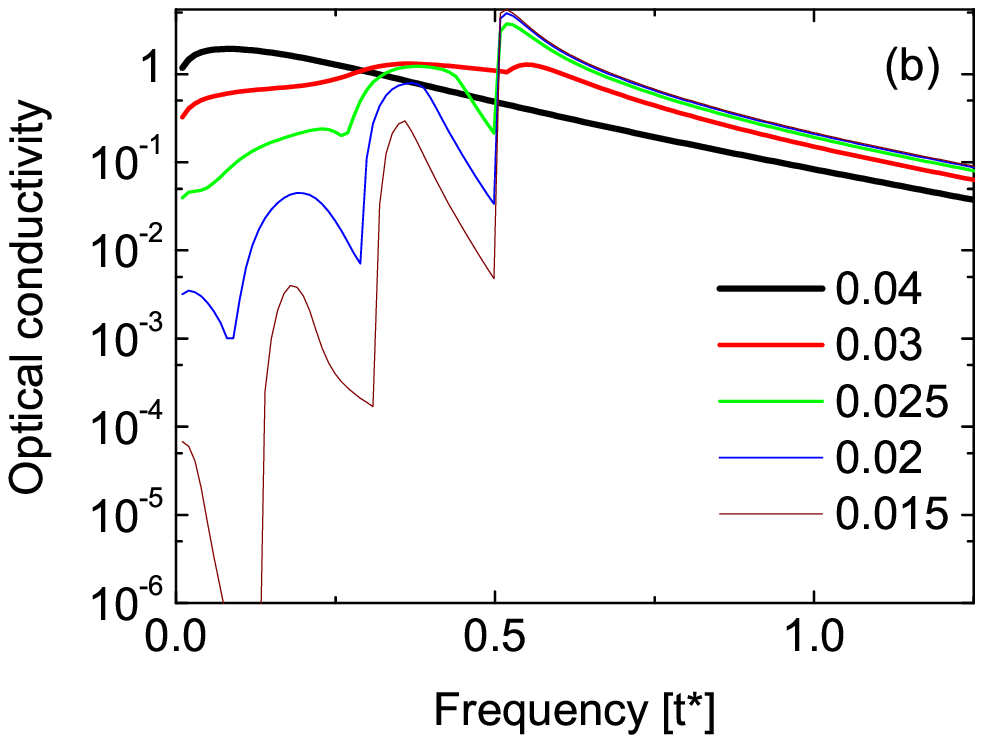}
  \caption{(Color online) Optical conductivity for $U=0.5$ and various temperatures.  
Panel (a) is a linear scale and panel (b) is a logarithmic scale. 
}\label{optgraf_u=0.5}
\end{figure}

We begin our discussion on transport properties in the ordered CDW phase by
examining the optical conductivity.
In Fig.~\ref{optgraf_u=0.5} we plot the temperature dependence of the optical 
conductivity for a dirty metal with $U=0.5$. At high temperatures we see the 
expected behavior for a dirty metal---namely, there is a peak at low energy and
a spread on the order of the metallic bandwidth.  The system does not have a low
energy fermi liquid peak, because it is not a fermi liquid. Below the critical 
temperature for CDW order, the shape of the optical conductivity changes
significantly. 
Note how the spectral weight is shifted
upward in frequency because the system is becoming an insulator at
low $T$.  In particular, a sharp peak 
develops at $\omega=U$ which corresponds to the interband transitions from the 
lower band at $\tilde\omega<-U/2$ to the upper band at $\tilde\omega>U/2$. 
We also see two additional peaks at lower frequencies. The higher of those peaks
corresponds to transitions from the lower band to the subgap states above 
the chemical potential and from the subgap states below the chemical potential 
to the upper band and the lower one corresponds to the transitions between the 
subgap states below and above the chemical potential. Both of these lower
energy peaks must vanish as $T\rightarrow 0$ because the subgap states
disappear continuously
at $T=0$.  Note that the frequency $\omega=U$ divides the spectra into 
two parts: to the right of this point the intensity of spectra increases 
as $T$ decreases and to the left of this point the intensity decreases  
as $T$ decreases which is similar to the isosbestic behavior of  
Mott insulators in the homogeneous phase, although we don't see the same kind of
isosbestic behavior in the ordered-phase optical conductivity here.

Results for $U=1$ have a similar structure to those for $U=0.5$, so we do 
not show them here.

\begin{figure}[htb]
 \includegraphics[width=0.4\textwidth]{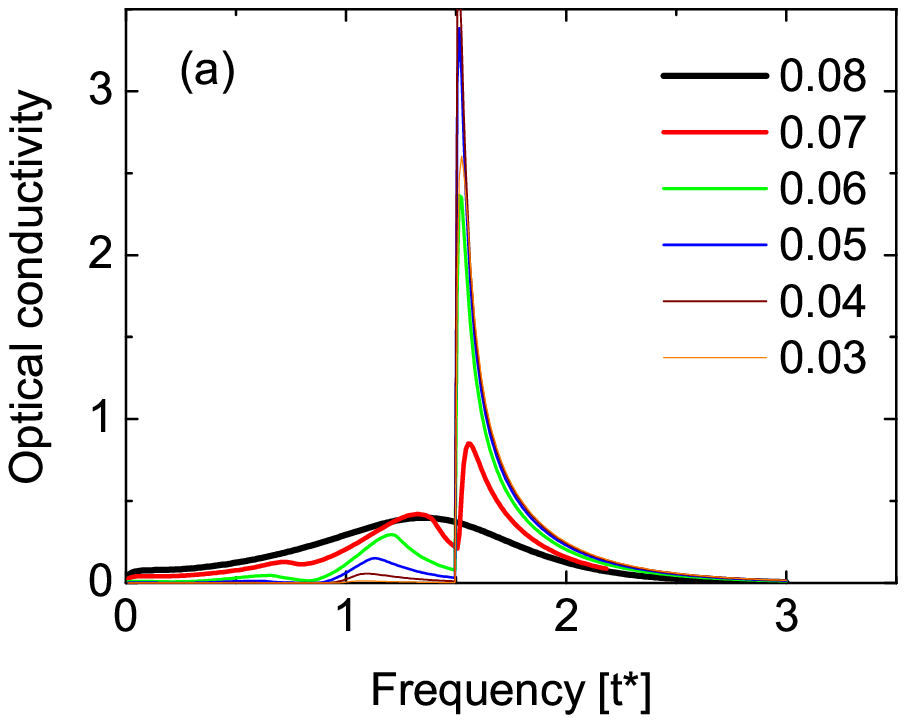}\\
 \includegraphics[width=0.4\textwidth]{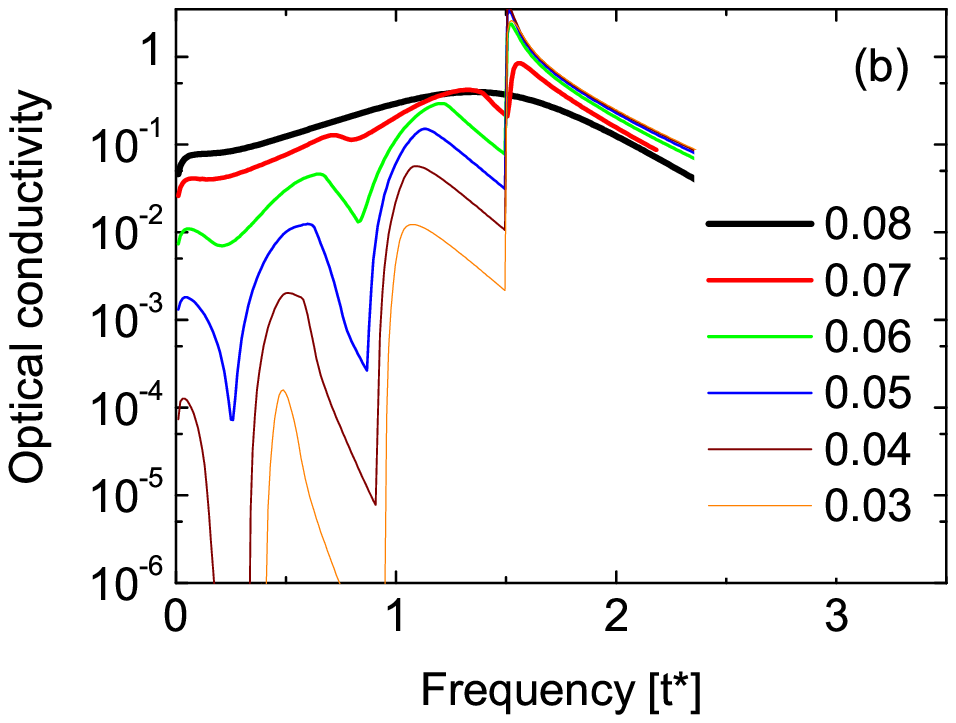}
  \caption{(Color online) Optical conductivity for $U=1.5$ and various temperatures.
Panel (a) is a linear scale and panel (b) is a logarithmic scale.
}\label{optgraf_u=1.5}
\end{figure}

We next plot the optical conductivity
for a near critical Mott insulator ($U=1.5$) in
Fig.~\ref{optgraf_u=1.5}. Here we see similar structure, with the 
peaks shifting to higher energy
as $T$ decreases, but the overall effect is not as large as in the metal,
because this system would be an insulator even if there was no CDW order.
Nevertheless, we still see the large peak develop with an
edge at $\omega=U$, and we see two low-energy peaks that have
strong temperature dependence due to the types of transitions
involving subgap states described above.

\begin{figure}[htb]
 \includegraphics[width=0.4\textwidth]{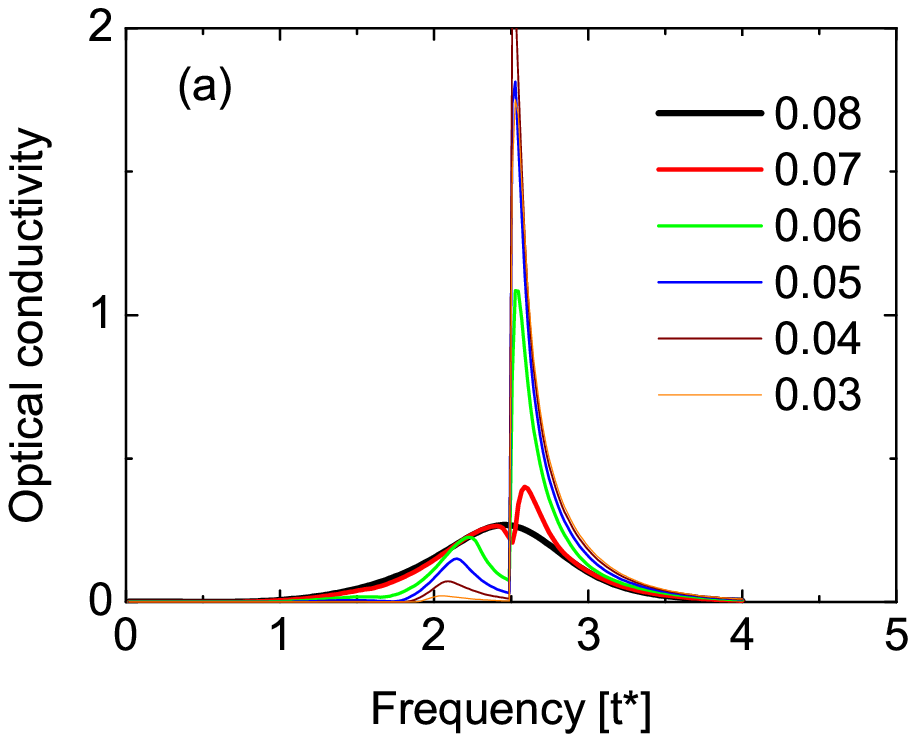}\\
 \includegraphics[width=0.4\textwidth]{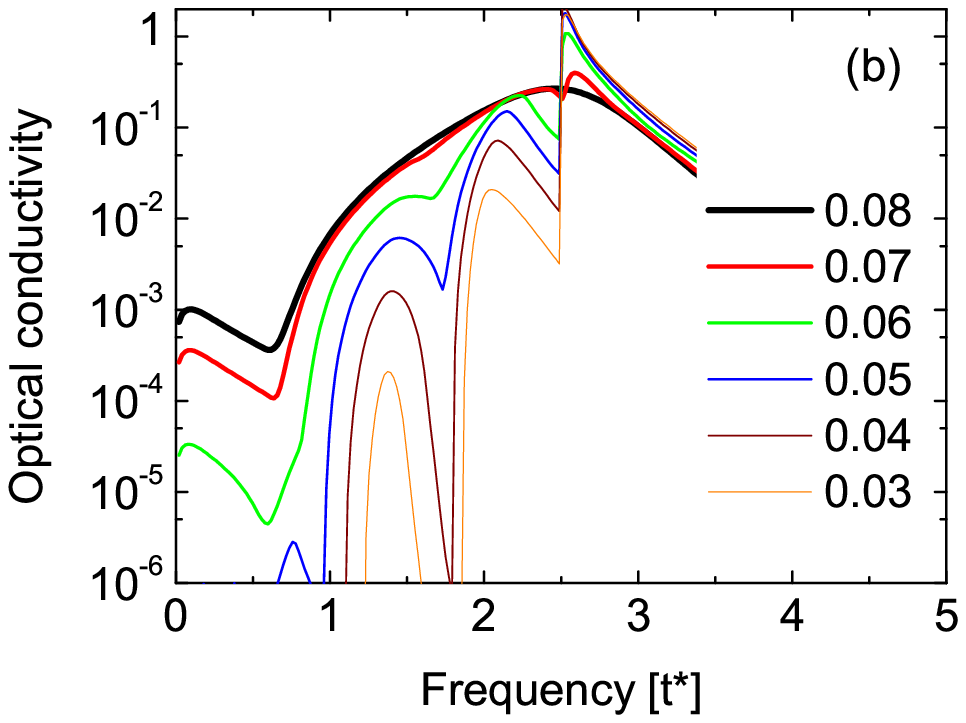}
  \caption{(Color online) Optical conductivity for $U=2.5$ and various temperatures.
Panel (a) is a linear scale and panel (b) is a logarithmic scale.
}\label{optgraf_u=2.5}
\end{figure}

Finally, we plot results for a moderate gap Mott insulator ($U=2.5$) in
Fig.~\ref{optgraf_u=2.5}.  The behavior here is essentially identical
to what we saw at smaller values of $U$, except the effects are smaller, 
because the subgap states are very small for frequencies below where the
Mott gap region extends, so the overall effects are somewhat reduced.  But
all of the qualitative behavior remains.

\begin{figure}[htb]
 \includegraphics[width=0.4\textwidth]{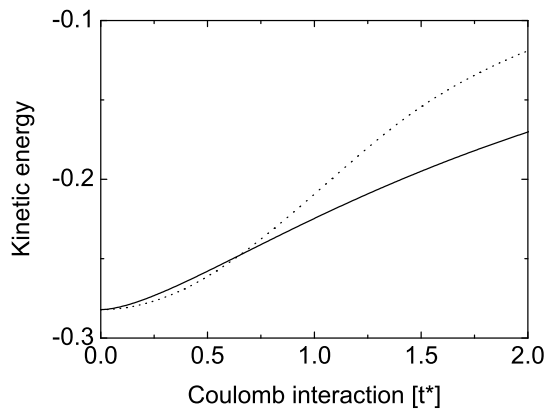}
 \caption{The average kinetic energy $K$ for different values of $U$ 
at $T=0$. The solid line corresponds to the CDW phase and the dotted line 
corresponds to the homogeneous solution. 
}
 \label{fig:KvsU}
\end{figure}

\begin{figure}[htb]
 \includegraphics[width=0.35\textwidth]{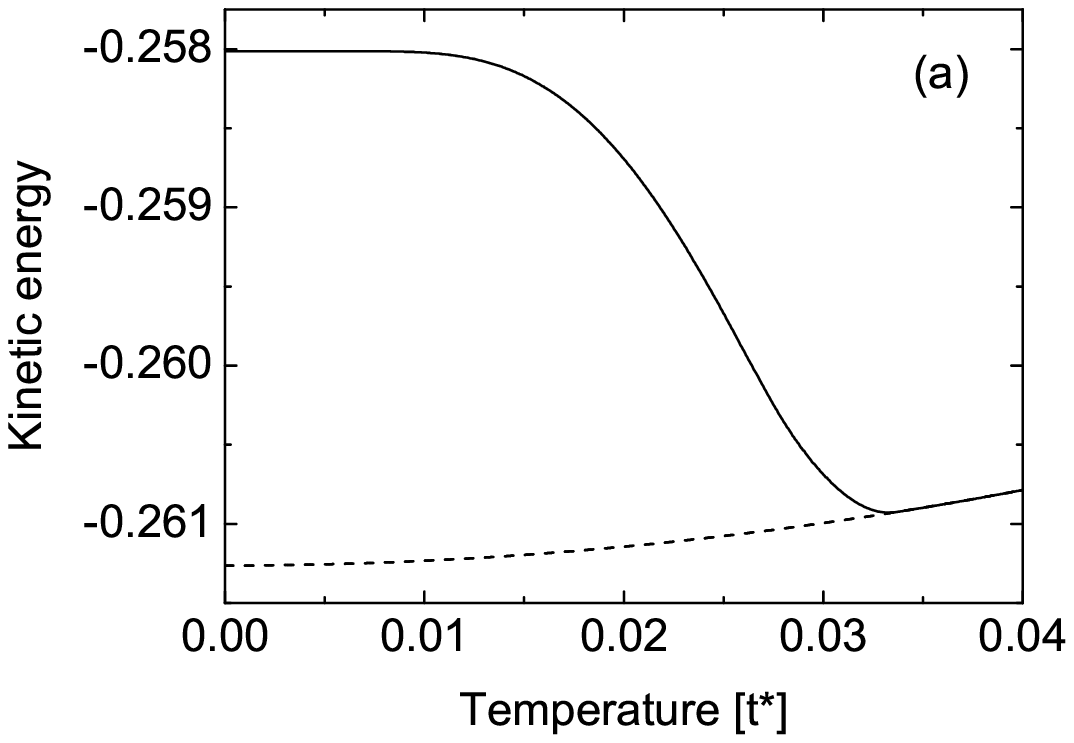}\\
 \includegraphics[width=0.35\textwidth]{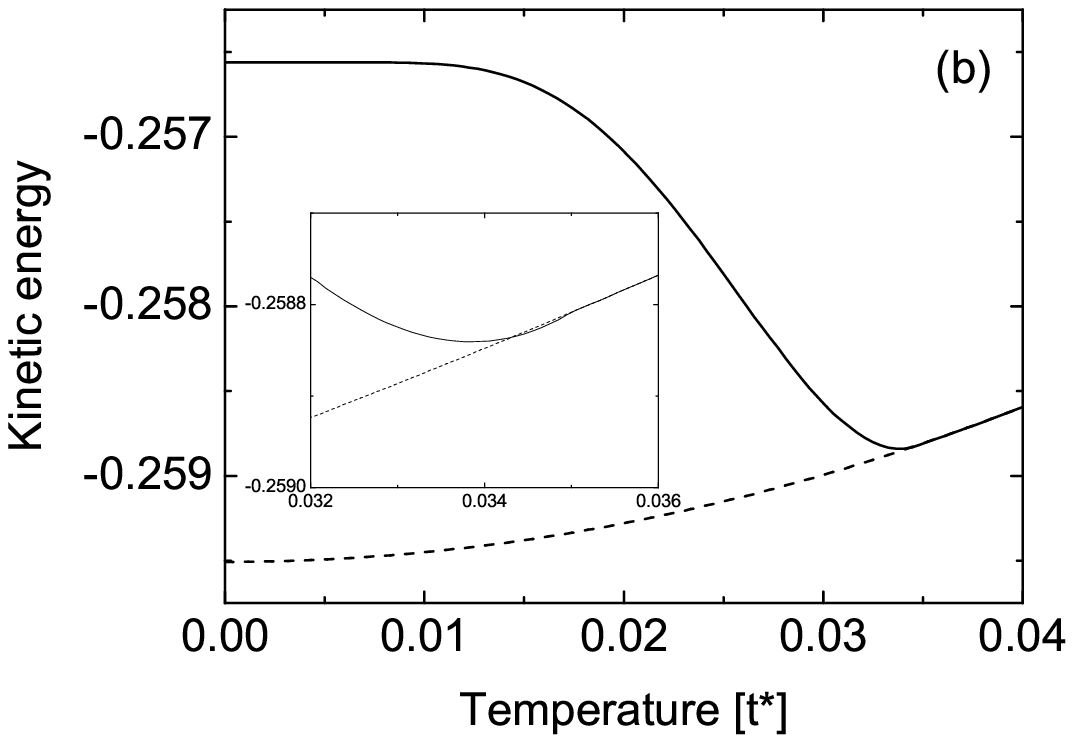}\\
 \includegraphics[width=0.35\textwidth]{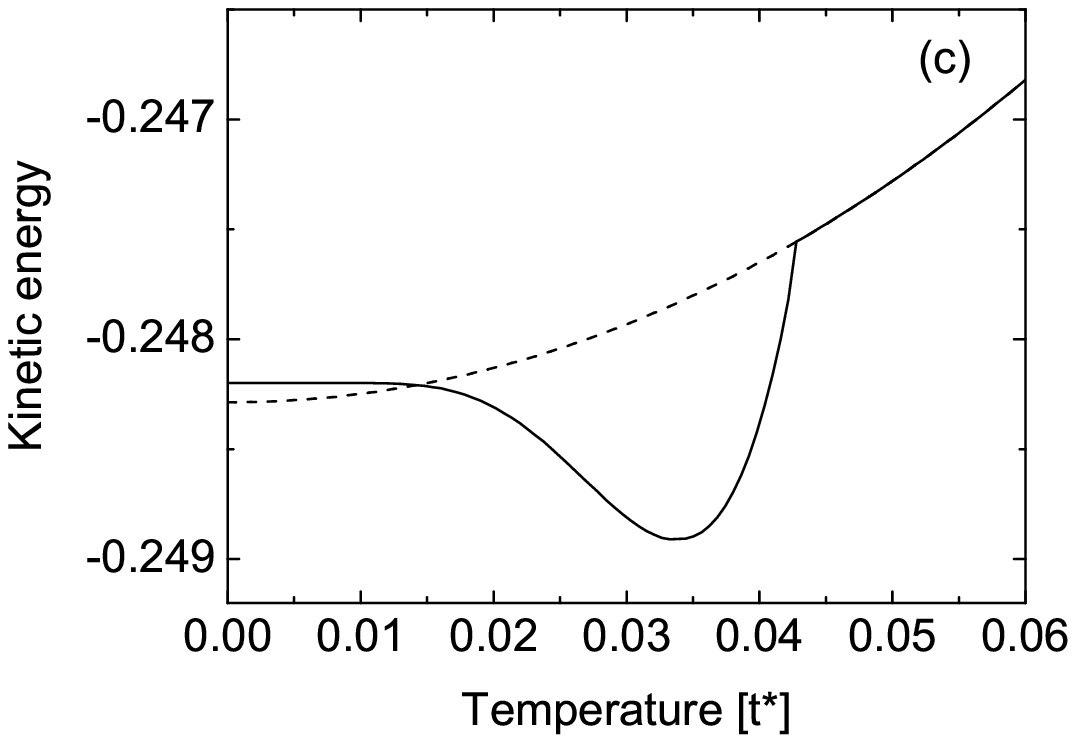}\\
 \includegraphics[width=0.35\textwidth]{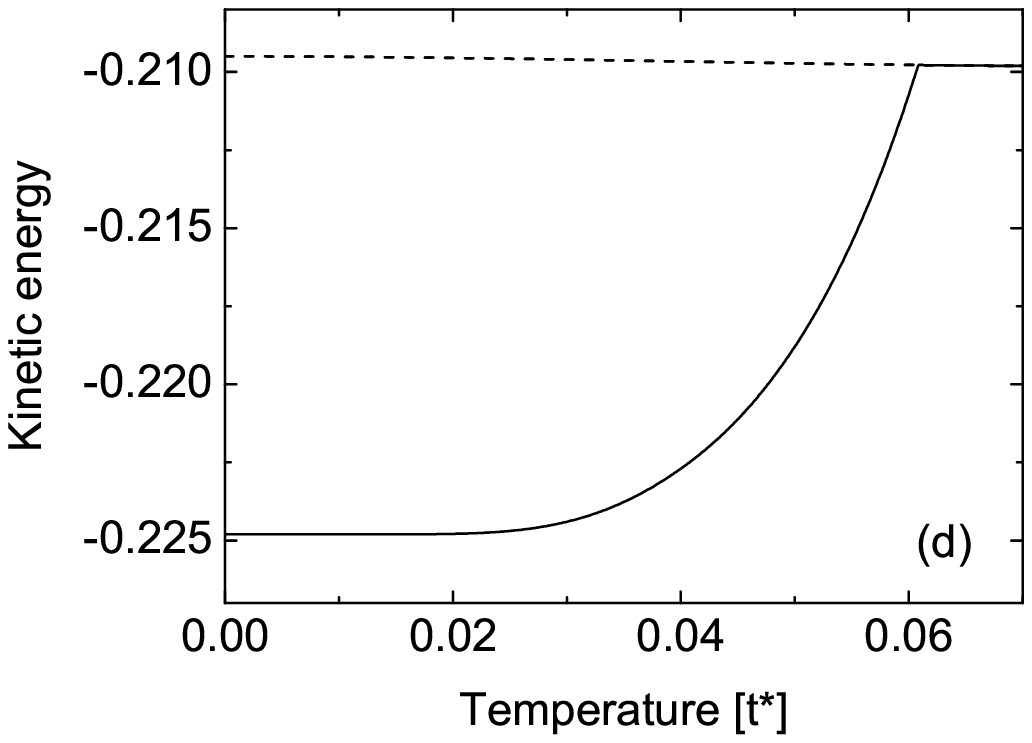}
 \caption{Temperature dependence of the average kinetic energy 
for different values of $U$: (a) $U=0.5$; (b) $U=0.52$; (c) $U=0.645$; 
(d) $U=1$. The solid line corresponds to the CDW phase and the dotted line 
corresponds to the homogeneous solution. }
 \label{fig:KvsT}
\end{figure}

In order to complete our discussion of the dynamical response, we now
describe the optical sum rule.
In general, the sum rule for the optical conductivity is
\begin{equation}
 \int_0^{\infty} d\omega \sigma(\omega) = - \pi K,
\end{equation}
where $K$ is the average kinetic energy (which is always 
nonpositive). In the CDW-ordered phase the average kinetic energy is equal to
\begin{align}
 K &= T\sum_m \frac1{2N}\sum_{\bm k} \epsilon_{\bm k} \left[ G_{m{\bm k}}^{AB} + G_{m{\bm k}}^{BA} \right]
\nonumber \\
   &= T\sum_m [\bar Z_m F_{zm} -1] = T\sum_m \lambda_m^A G_m^A = T\sum_m \lambda_m^B G_m^B
\nonumber \\
   &= -\frac{1}{\pi}\int d\omega f(\omega) \Img \lambda^a(\omega) G^a(\omega), \quad a=A,B,
\end{align}
and at $T=0$, when $\Sigma^A=U$ and $\Sigma^B=0$, we immediately find
\begin{equation}
 K=-\frac12\int d\epsilon\rho(\epsilon)\frac{\epsilon^2}{\sqrt{\frac{U^2}{4}+\epsilon^2}}.
\end{equation}
In Fig.~\ref{fig:KvsU}, we plot the average kinetic energy 
both for the CDW and homogeneous solutions for different values of $U$ at 
$T=0$. For small values of $U$ ($U<0.648$), we observe  the 
anticipated behavior 
that the average kinetic energy increases faster in the 
ordered phase than in the homogeneous phase.  This is anticipated because the
homogeneous phase has, on average, some neighboring sites with no localized 
electrons, implying hopping is easier than in the ordered phase, where 
every hop involves a change in energy by $U$ at $T=0$ because the order
parameter is uniform on each sublattice.  Since it is more difficult to hop in 
the ordered phase, the kinetic energy increases relative to the homogeneous
phase.  For large values of $U$ we find anomalous behavior, where the average
kinetic energy is more negative in the ordered phase. There is no simple picture
to explain how this occurs.  In the homogeneous phase, as $U$ increases, it
becomes more difficult to hop because the doubly occupied states are being
projected out of the system.  This implies the average kinetic energy increases
in the homogeneous phase, but it does so faster than in the ordered phase.
The subtle details of how
the average kinetic energy evolves with temperature are shown in 
Fig.~\ref{fig:KvsT}.
The anomalous behavior for the temperature dependence of the
average kinetic energy occurs for a finite range of $T$ when $U> 0.52$. 
This is the ``critical'' $U$
value where the DOS in the normal state changes its curvature from being
negative at the 
chemical potential, as expected for a conventional metal, to positive in what
is sometimes called an anomalous metal. In the region
$0.52<U<0.648$, the normal state DOS starts to develop a dip at the chemical
potential, and for a finite temperature range, the anomalous behavior in the
average kinetic energy occurs only for low temperatures. As $U$ is increased
further, we see the anomalous behavior occur for all $T$.
These results show that the spectral weight in the CDW phase shows a modest
decrease for small $U$ and a dramatic increase for large $U$ at $T=0$!
This is somewhat unexpected, since the behavior is different than what
is seen in say a BCS superconductor, where the gap formation reduces
spectral weight at high frequencies, but the lost weight is restored
in a zero frequency Drude peak.  For the CDW ordered phase, no zero
frequency delta function appears.  The spectral weight loss is small for
small $U$, but the gain can become significant for large $U$.

\begin{figure*}[htb]
\includegraphics[width=0.35\textwidth]{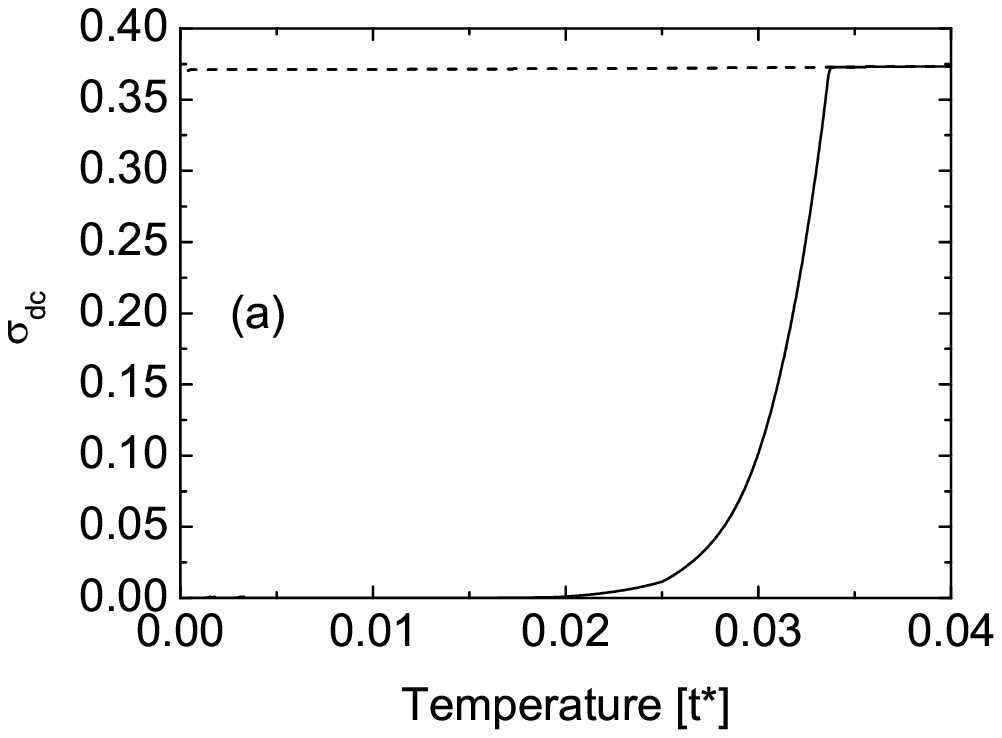}
\qquad\includegraphics[width=0.35\textwidth]{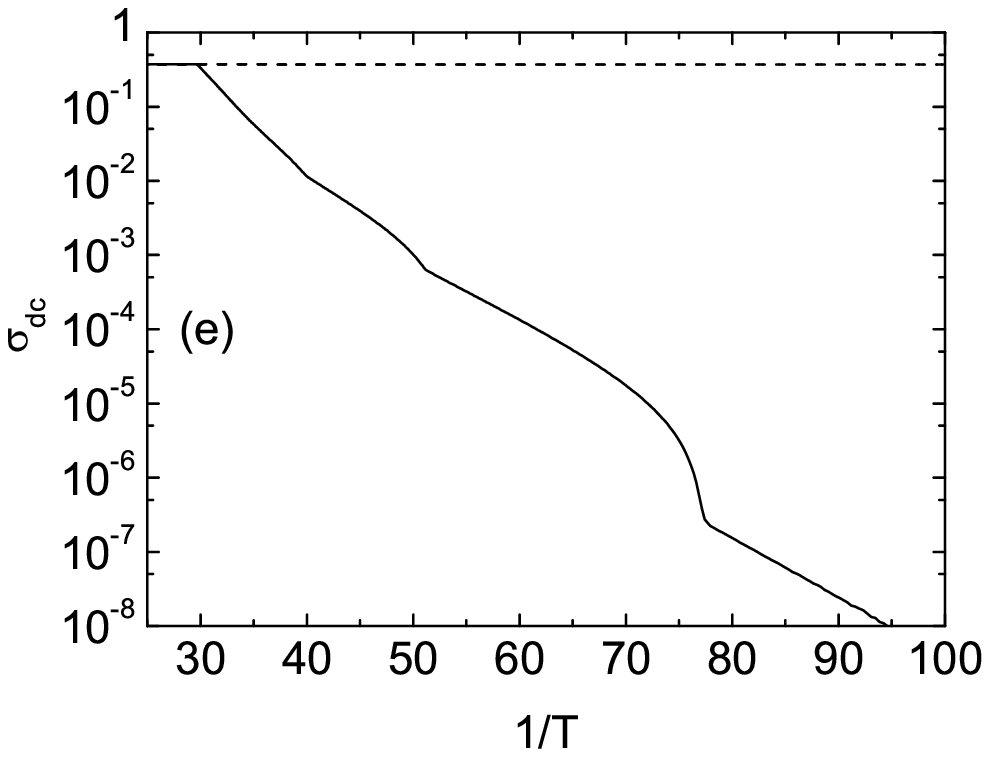}\\
\includegraphics[width=0.35\textwidth]{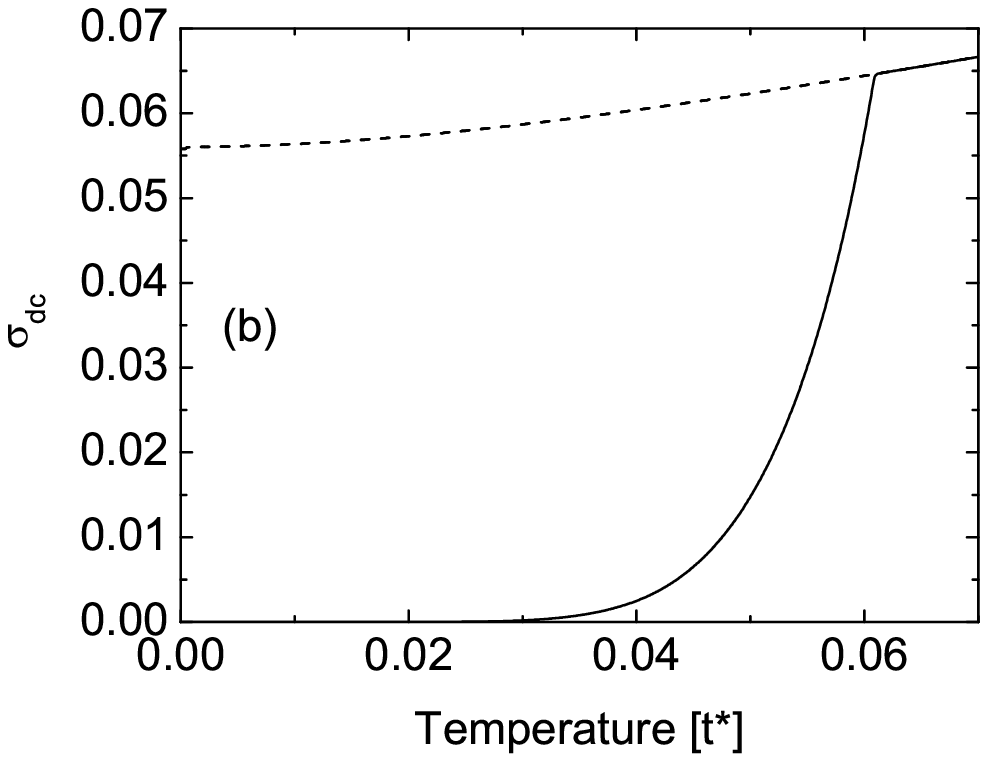}
\qquad\includegraphics[width=0.35\textwidth]{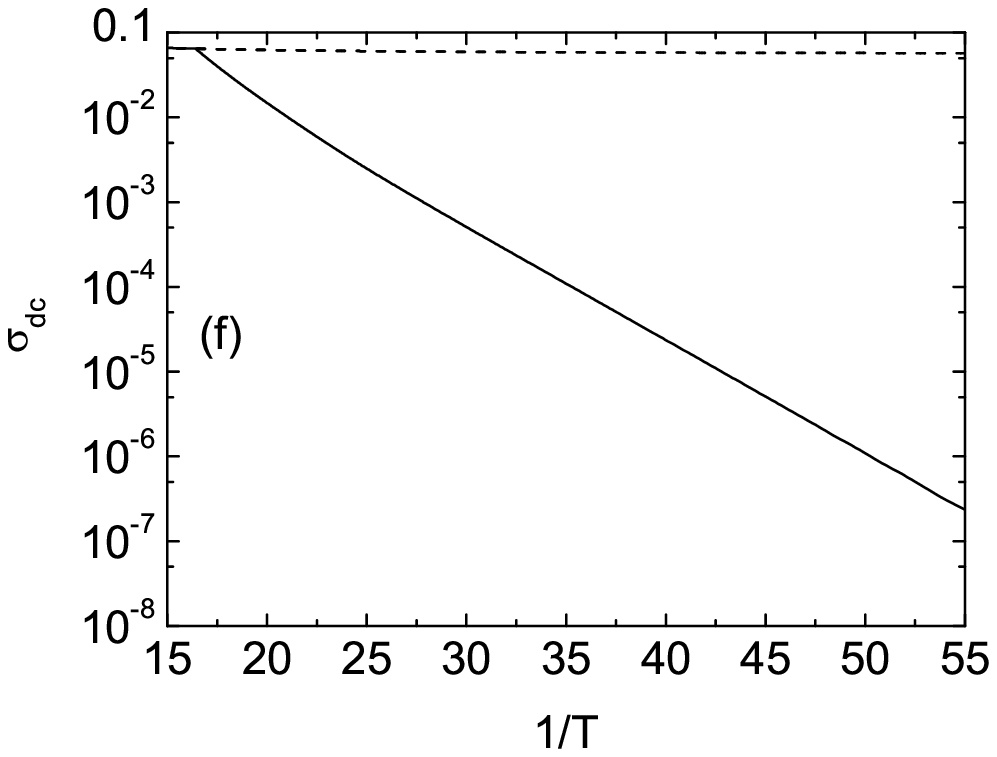}\\
\includegraphics[width=0.35\textwidth]{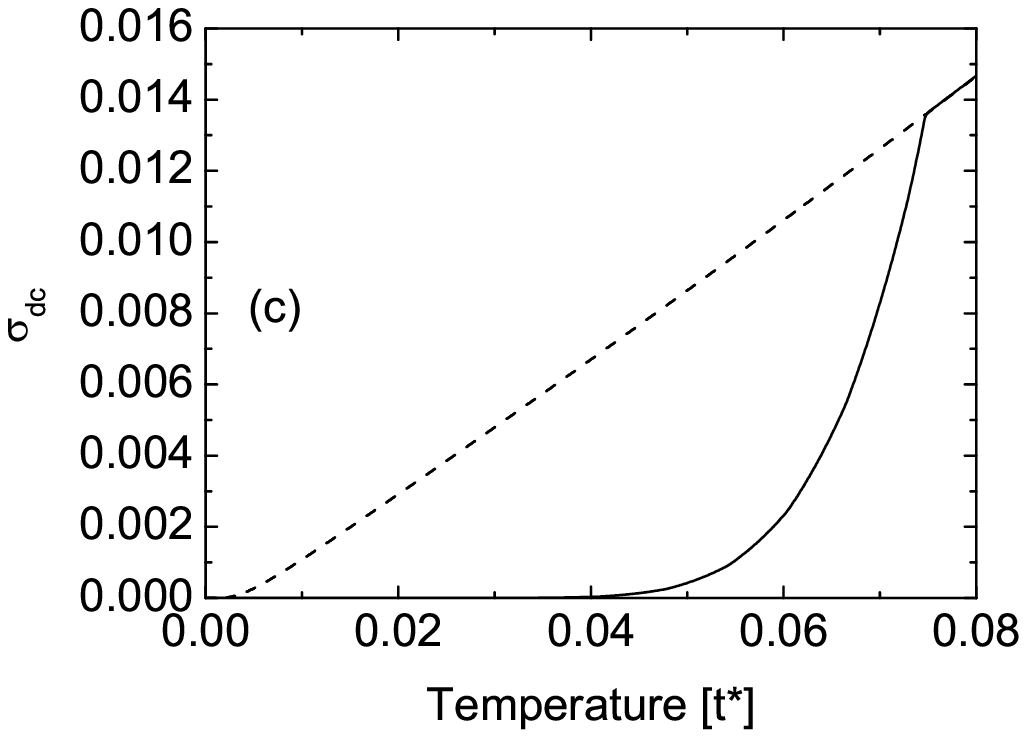}
\qquad\includegraphics[width=0.35\textwidth]{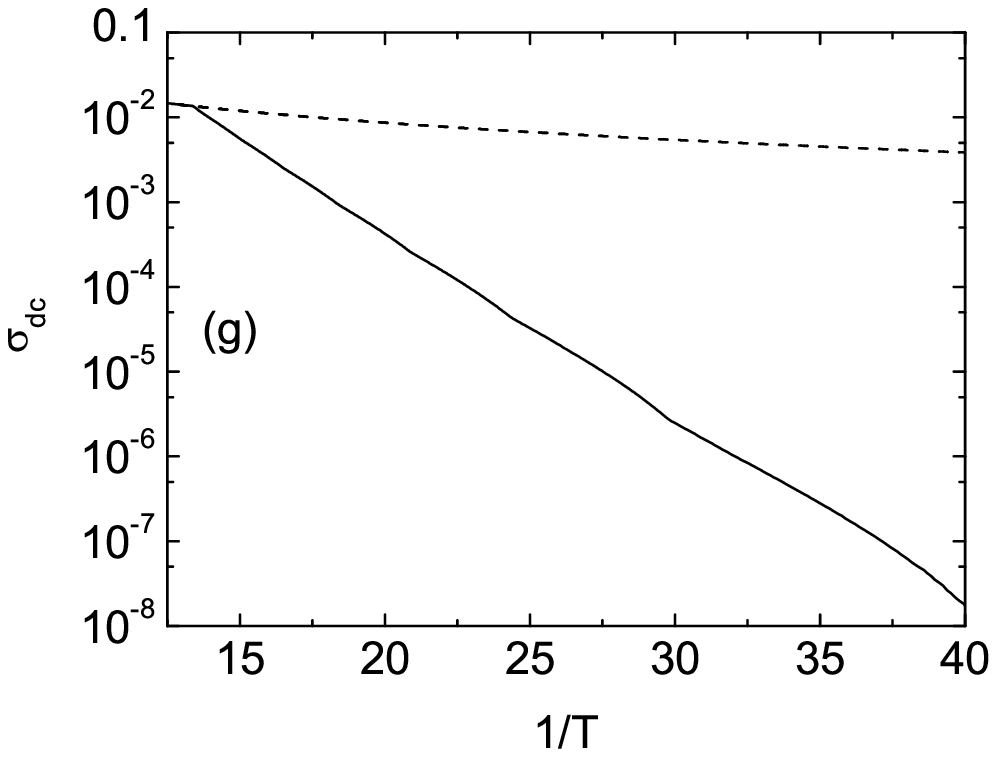}\\
\includegraphics[width=0.35\textwidth]{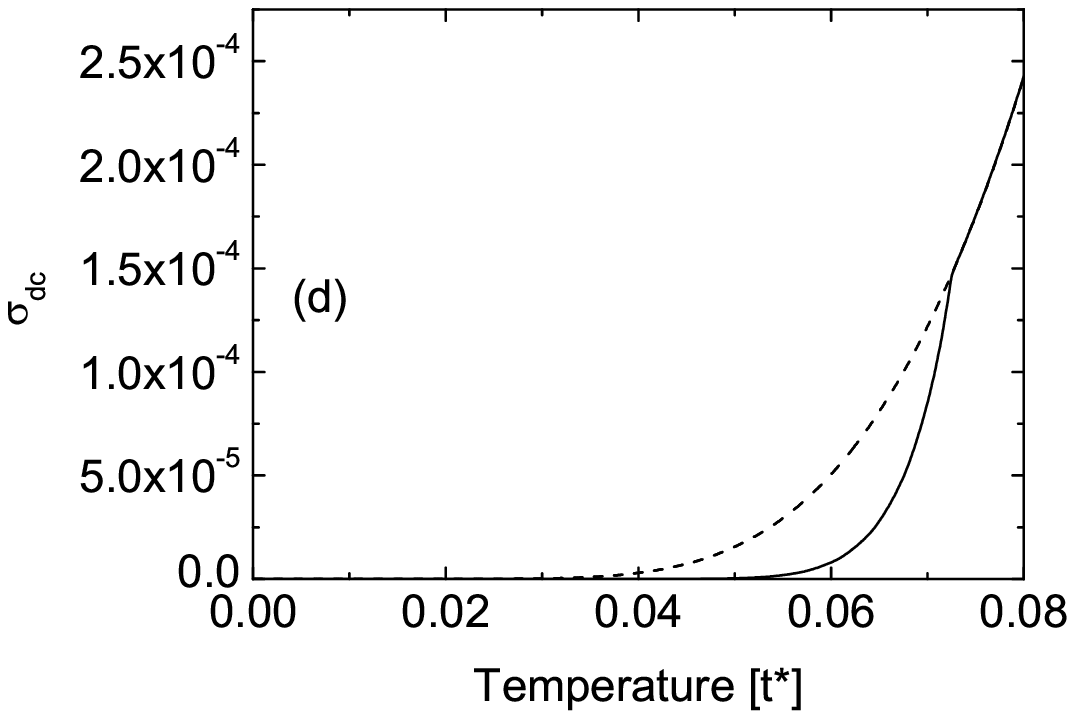}
\qquad\includegraphics[width=0.35\textwidth]{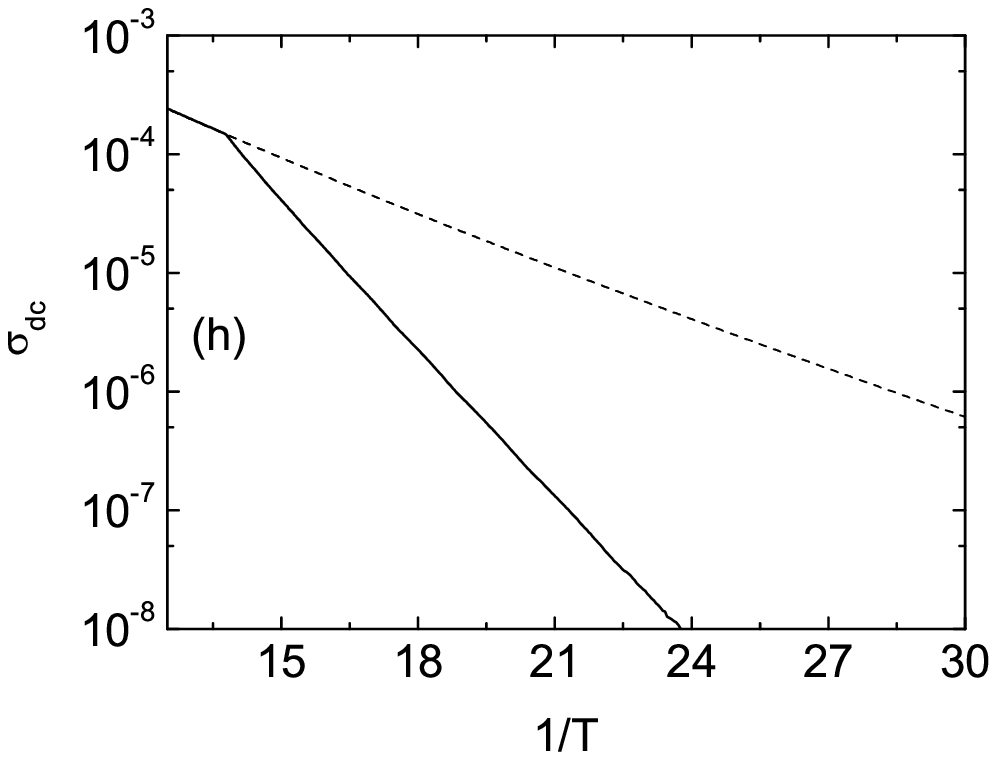}
\caption{dc conductivity for (a) $U=0.5$
($T_{c}\approx0.034$), (b) $U=1$ ($T_{c}\approx 0.0615$), (c) $U=1.5$ ($T_{c}\approx0.075$) and (d)
$U=2.5$ ($T_{c}\approx0.072$) as a function of temperature. The
solid line denotes the CDW ordered phase and the dashed
line denotes the homogeneous one. Results are presented on a linear scale (left) and logarithm of dc conductivity vs inverse temperature (right).} \label{optconst}
\end{figure*}

\begin{figure*}[htb]
\includegraphics[width=0.35\textwidth]{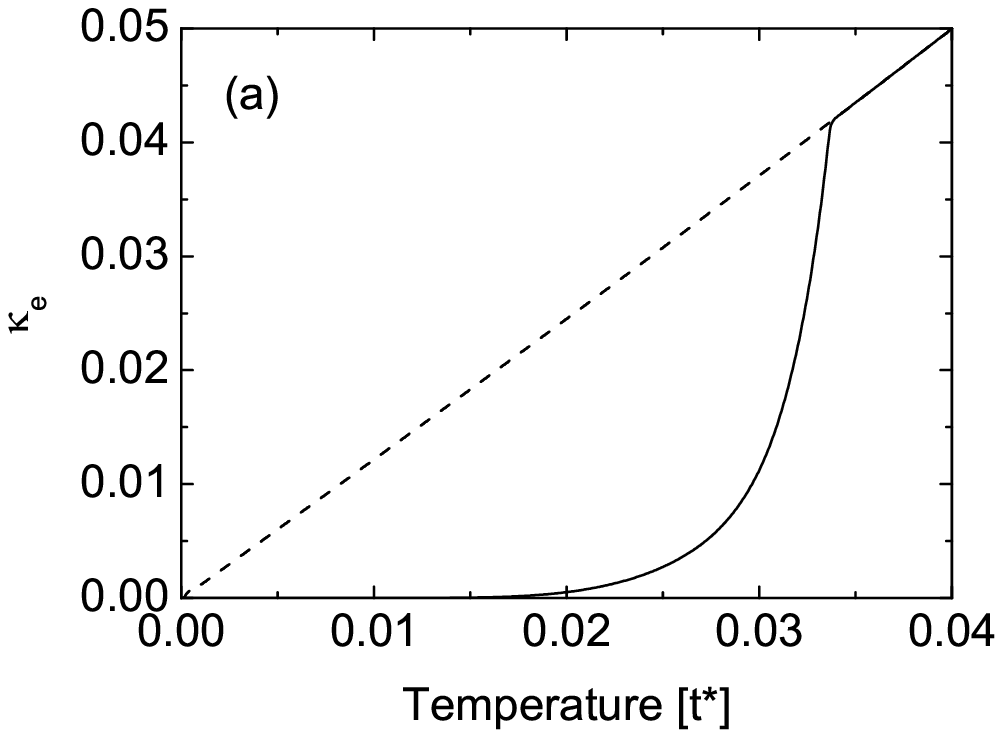}
\qquad\includegraphics[width=0.35\textwidth]{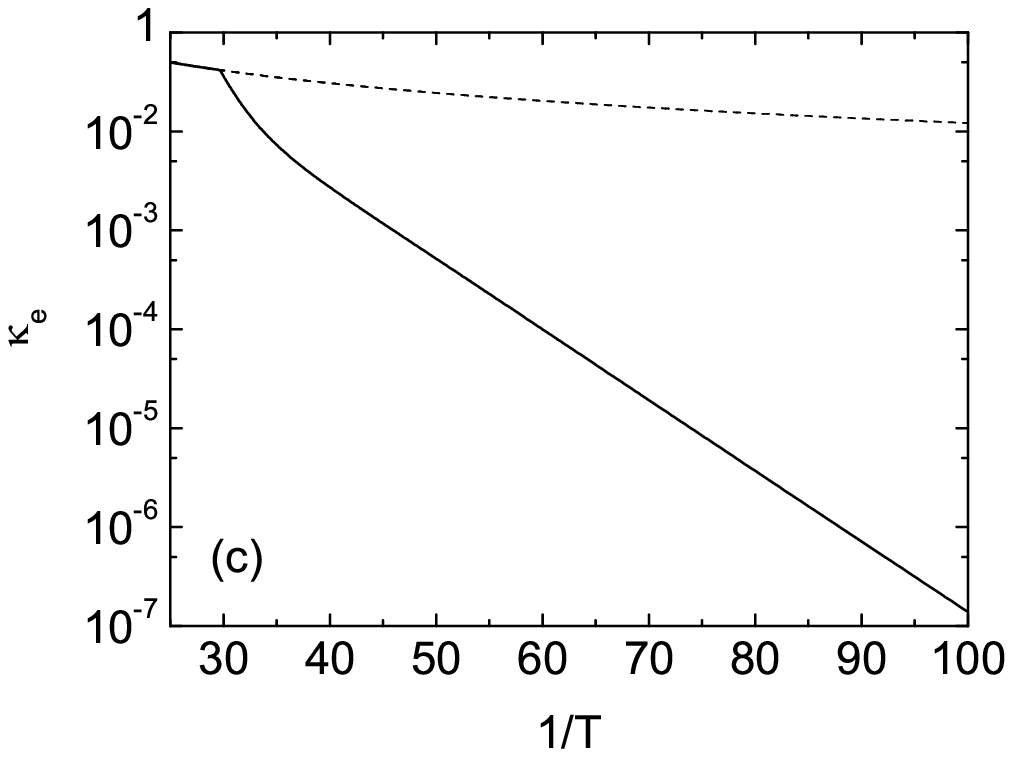}\\
%\includegraphics[width=0.35\textwidth]{therm_cond10}
%\qquad\includegraphics[width=0.35\textwidth]{lnK_invT_u10}\\
%\includegraphics[width=0.35\textwidth]{therm_cond15}
%\qquad\includegraphics[width=0.35\textwidth]{lnK_invT_u15}\\
\includegraphics[width=0.35\textwidth]{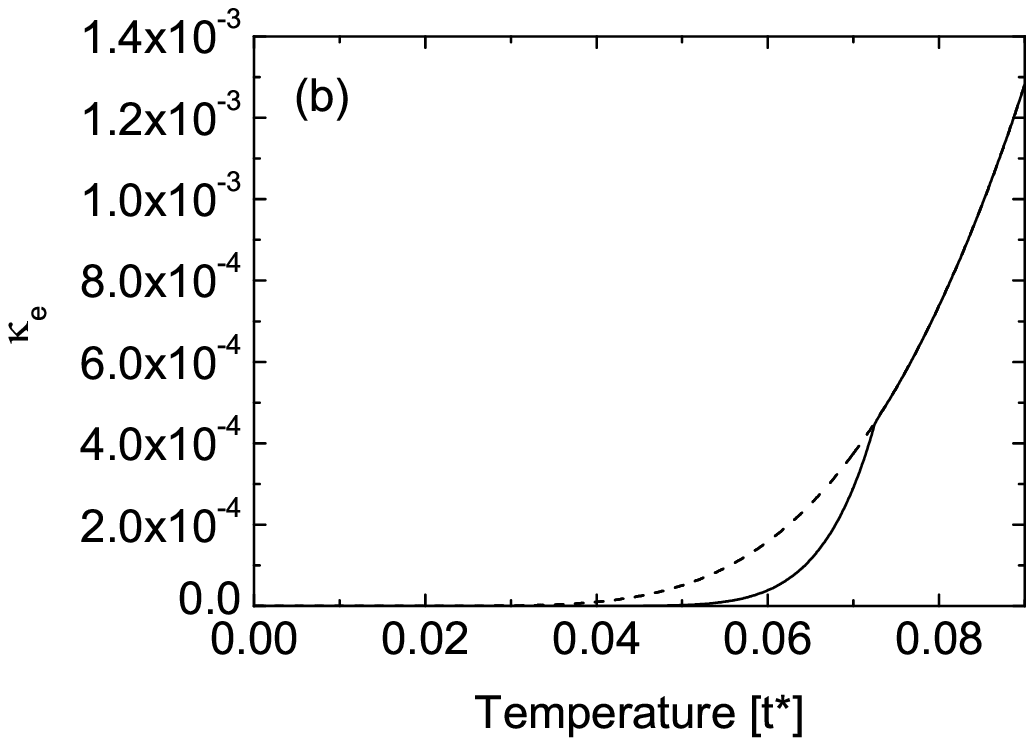}
\qquad\includegraphics[width=0.35\textwidth]{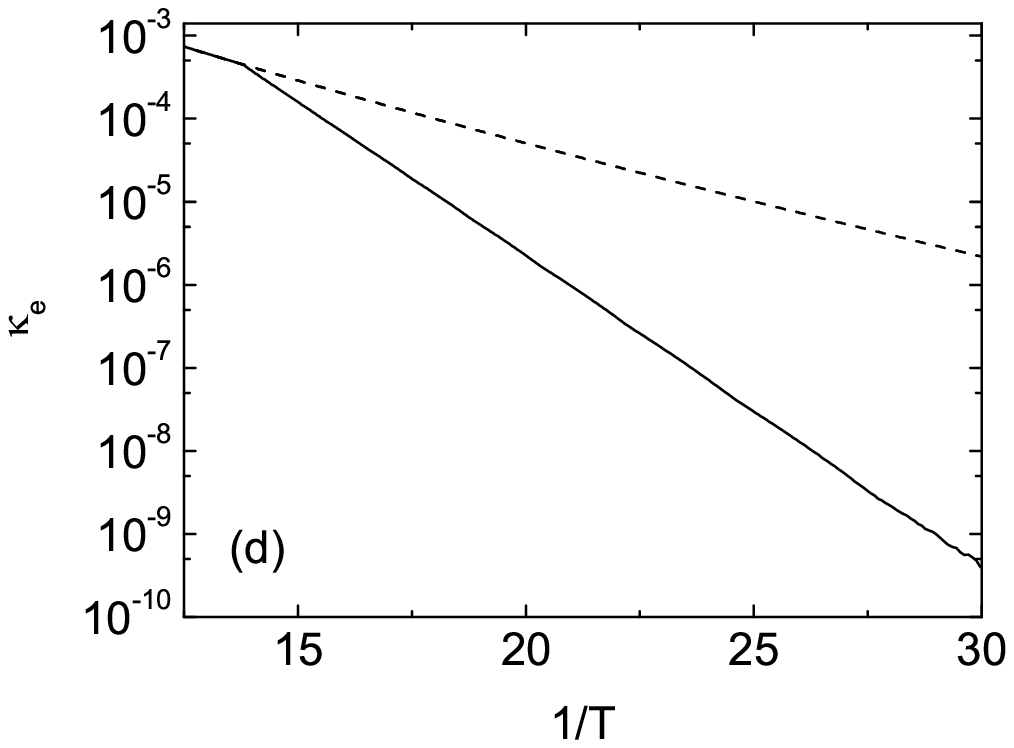}
\caption{Electronic contribution to the thermal conductivity for
(a) $U=0.5$ ($T_{c}\approx0.034$), 
and (b) $U=2.5$ ($T_{c}\approx0.072$)
as a function of temperature. The solid line denotes the CDW
ordered phase and the dashed line denotes the homogeneous
one. Results are presented on a linear scale (left) and logarithm of thermal conductivity vs inverse temperature (right).} \label{thermcond}
\end{figure*}

Next we examine the dc transport.
The temperature dependence of the dc and thermal conductivity are plotted in 
Figs.~\ref{optconst} and \ref{thermcond}, respectively, where we plot both 
the CDW solution and the homogeneous solution extrapolated into the CDW region. 
At low temperatures, due to the factor $-df(\omega)/d\omega$, the 
main contributions to the dc transport come from the narrow region of 
width $4T$ around the chemical potential (the so-called fermi window). For 
the Falicov-Kimball model at half filling in the homogeneous phase ($T>T_c$) 
the DOS, Green's functions and self-energies do not depend on temperature 
and, as a result, the temperature dependence of the dc transport is determined
solely by the shape of the relaxation time in Eq.~(\ref{RT_def}) close to the 
chemical potential. For small $U$ values the relaxation time 
$\tau(\omega)$ is flat [Fig.~\ref{fig:rel_t}~(a)] and, as a result, the dc conductivity for the 
homogeneous phase is essentially a constant for low $T$. 
At $U=\sqrt2$ the Mott insulator forms.  For
larger $U$ values, one might expect to see exponentially activated transport,
but that does not occur on the hypercubic lattice, because the system only
possesses a pseudogap.  Even though the DOS exponentially decreases in the 
gap region, the lifetime of the excitations is exponentially long, and
$\tau(\omega)$ behaves like $\omega^4$ for low energies\cite{demchenko}.  
This produces
a quartic dependence of the dc conductivity on $T$, and a higher power
law for the thermal conductivity. 

In the CDW phase ($T<T_c$), the CDW gap is filled by subgap states  at finite
$T$, which lead to a less severe modification of the exponentially activated 
transport at low $T$. But it is only the subgap states within the fermi
window that affect the transport, so the modification is not quite as
severe as one might have naively guessed.
Note the small wiggles in the solid lines at low $T$.  These occur due to the
evolution of the subgap states.  The $T$ dependence of the dc transport
always shows a marked kink at $T_c$ with the conductivities sharply suppressed
as the CDW gap forms.  In the Mott insulator, the transport changes from power
law in $T$ to exponential activation (suitably modified by the subgap states).
The thermal conductivity displays similar features, as shown in Fig.~\ref{thermcond}.

\section{Conclusions}

In this work we have developed the formalism to calculate transport properties
of CDW-ordered phases within DMFT.  Since the dc charge and heat transport
and the optical conductivity continue to have no vertex corrections, even
in the ordered phase, the calculations reduce to a careful evaluation of the
bare Feynman diagrams with a sublattice index introduced by the order.

As the system orders into a CDW state, the DOS develops a gap with a sharp
singularity in the DOS at the band edge when $T=0$.  The gap at $T=0$ is
always equal to $U$.  As the temperature increases, but still below $T_c$,
we see a significant development and evolution of subgap states within
the gap region.  This gap region where subgap states develop, appears to
lie within the extent of the normal-state DOS---in other words, in the
Mott insulator, we do not see subgap states develop within the region that
corresponds to the Mott gap in the normal state.  We verify the accuracy of
the DOS calculations by calculating the zeroth, first, and second moment
of the local DOS on each sublattice and we find they agree with exact results
to essentially machine accuracy.

The optical conductivity has a significant rearrangement of states within 
the ordered phase, which can be understood by examining the different kinds of
processes that take place within an optical transition---namely that we move 
from an occupied to an unoccupied state. Because there are many different
bands that are present at finite $T$, this leads to significant structure in the
optical conductivity.   In particular, the singularity in the DOS leads to a 
large asymmetric peak centered around $U$ in the response function.  The
total spectral weight is governed by the average kinetic energy due to the
optical sum rule.  While a naive expectation would say the average kinetic
energy increases when the ordering is turned on (\textit{i. e.}, it becomes less
negative with a smaller magnitude) because the ordering blocks hopping between
the sublattices,  we find that is true only for small $U$. For small $U$
the kinetic energy shows a modest increase, so some spectral weight is lost due
to the ordering.  For larger $U$ the kinetic energy shows a significant
reduction (\textit{i. e.}, the magnitude increases as the average kinetic energy
becomes more negative) so the spectral weight increases when the
ordered phase is entered, and that increase can become quite substantial
as $U$ becomes large. 

Finally, we also examined the dc transport.  Since we are at half
filling, one can show the thermopower vanishes due to particle-hole symmetry
even in the presence of CDW order.  Hence we can only examine the charge
and heat conductivities. We find that the CDW order suppresses both of these,
but because of the subgap states and their complicated evolution with
temperature, the dc response does not obey any simple functional
form at low $T$.  Instead, we often see significant wiggles in the 
conductivities.  In the Mott-insulating phase, the conductivity should go from
a power-law-like behavior to exponential activation.  We see such a trend
start to develop, but we cannot accurately quantify this because we cannot go
down far enough in temperature in the CDW phase before we run into issues with
accuracy of the calculations.

This work shows that there is rich and interesting behavior that occurs in
the transport as CDW order sets in. In future work, we will examine 
Raman scattering, where vertex correction effects are present and 
inelastic X-ray scattering, where interesting phenomena is likely to
occur when the photon transfers momentum equal to the ordering wavevector.

\acknowledgments

This publication is based on work supported by Award No. UKP2-2697-LV-06
of the U.S. Civilian Research and Development Foundation.  
JKF was also supported
by the National Science Foundation under Grant No. DMR-0705266.  We would
like to acknowledge useful discussions with Tom Devereaux.

\end{document}